\documentclass[pra,letterpaper,twocolumn,superscriptaddress,floatfix]{revtex4}
\usepackage{graphicx,psfrag,amsmath,amssymb,amsfonts,bbm,latexsym,color,dcolumn,bm}

\begin{document}
	
\title{Generation of Non Maxwell-Boltzmann energy distributions during \\  thermalization of interacting baths}

\author{Roberto Onofrio}

\affiliation{\mbox{Department of Physics and Astronomy, Dartmouth College, 6127 Wilder Laboratory, Hanover, NH 03755, USA}}

\author{Bala Sundaram}

\affiliation{\mbox{Department of Physics, University of Massachusetts, Boston, MA 02125, USA}}

\begin{abstract}
We report on  numerical experiments of thermalization between two reservoirs in which interspecies interactions are mediated 
via a Hamiltonian with a Gaussian dependence upon the distance between the particles.  By using relatively high statistics, we are 
able to study deviations from the initially imprinted Maxwell-Boltzmann energy distributions as a function of the range and strength of the 
interaction, as well as the parameters of the two baths such as masses, trapping frequencies, that strongly affect the spatial overlap of the two species. 
At large interaction strengths and for any range, deviations are attributable to the sudden conversion of latent interaction energy into 
kinetic energy of the particles. In addition, long range interactions generate deviations even for interaction strengths in the weak regime, and 
non-uniform confinement introduces a further source of deviation from the Maxwell-Boltzmann energy distribution. This latter results in transient two temperature 
Maxwell-Boltzmann distributions suggestive of an inverse cascading route from inter-particle dynamics to thermalization.
In the weakly-interacting regime, a simple expression for the equilibrium temperature is also obtained and discussed.
These findings may be of relevance in the dynamics of exothermic chemical reactions, in the physics of ultracold atomic mixtures, and in 
self-sustained burning plasmas.
\end{abstract}

\maketitle

\section{Introduction}

The study of statistical properties of interacting many-body systems plays a crucial role in classical and quantum physics, 
from soft condensed matter and plasma physics to mesoscopic quantum systems. In the prototypical case of systems 
with a Maxwell-Boltzmann (MB) energy distribution, the statistical properties are summarized by assigning probabilities 
for the occurrence of energy as proportional to the Boltzmann factor, exponentially suppressing high energy states 
as $p(E) \propto \exp(-\beta E)$, with $\beta$ as the only free parameter of the distribution, the inverse temperature.

Particular intriguing is the discussion of the approach to thermal equilibrium characteristic of an MB energy distribution, the 
subject of kinetic theory since the foundation of statistical mechanics \cite{Maxwell,Boltzmann,Cercignani,Kremer}.
Given two statistical systems each described by an MB distribution, with different inverse temperatures, it 
is important to understand if, when interacting, they exchange energy in such a way as  to reach a stationary state, and if this 
stationary state is still described by an MB distribution. This problem has been tackled with both analytical 
and numerical tools. For instance, in a sequence of papers analytical conditions have been discussed 
for maintaining MB states at all times \cite{Rubin1,Rubin2,Rubin3,Montroll,Andersen}. In recent numerical simulations, 
starting from non-MB energy distributions for systems of anharmonic oscillators connected by few harmonic 
oscillators, thermalization between two baths to a final MB distribution was achieved \cite{BagchiTsallis}. 

In prior publications, we have explored the thermalization dynamics of two ensembles of classical particles, 
each with initial conditions drawn from MB energy distributions, and with interspecies interactions  
characterized by two free parameters, the interaction strength $\gamma$ (representing the maximum energy exchange 
in a single interspecies interaction) and the interaction range $\lambda$. These two parameters appear in the interaction 
Hamiltonian such that the contribution due to two generic particles, one from each of the two baths with coordinates $q_n$ and $Q_m$ 
will be
\begin{equation}
H_{\mathrm{int}}(q_n, Q_m)= \gamma~ e^{-(Q_m-q_n)^2/\lambda^2},
\label{Hint}
\end{equation}
a choice inspired by a microscopic model for a meter in quantum measurement theory \cite{Mensky} and already adopted in our previous contributions.
More specifically, we have discussed the dynamics of the thermalization process \cite{OnoSun1}, the presence of anomalous scaling 
of the interaction energy with number of particles in a balanced mixture \cite{JauOnoSun1}, the interplay of thermalization 
and nonlinearities included in the trapping potential \cite{JauOnoSun2}, the dependence of the anomalous scaling upon 
the interaction strength $\gamma$ \cite{JauOnoSun3}, and upon the interaction range $\lambda$ \cite{OnoSun2}. 

In this study, we consider in detail the dependences of the energy distributions during the thermalization dynamics, especially at short times, 
on the two parameters, $\lambda$ and $\gamma$, the trapping frequencies and masses, and the number of particles in each bath.  
In particular, we show the importance of the interaction range $\lambda$ in controlling deviations from MB distributions. 
This is relevant for at least two reasons. The presence of non-MB energy distributions has already 
been reported in the so-called Hamiltonian Mean-Field (HMF) model \cite{Konishi,Ruffo}, which however is not representative of classical 
dynamical systems for which most of statistical mechanics has been developed. Additionally, the presence of non-MB energy distributions, 
and most notably of `hard' high-energy tails, may play an important role in accelerating fusion processes in thermonuclear plasmas \cite{Majumdar,Onofrio,Xie1,Kong,Squarer}. 

The paper is organized as follows. In Section II we first define the setting for the interaction between the two baths and the related simulations, 
and discuss the results for the long-time dynamics. In Section III we instead focus on transient dynamics and observe the dependence of deviations from MB
energy distributions and their dependence upon $\lambda$ and $\gamma$. Section IV is devoted to the analysis of the dependence of the transient upon mass and angular frequency 
of the baths.  Slower thermalization results when there is mismatch between these parameters, due both to the intrinsic inefficiency of two particle 
interactions as well to the diminished overlap between the two species in the respective harmonic potentials. Section V contains analytical results for the equilibrium 
temperature of the two baths in the weakly-interacting regime. The connection of our results to various subfields in which they may be relevant, such as 
plasma physics, cold atoms, and chemical reaction dynamics, is briefly discussed in the conclusions.

\section{General considerations on thermalization}

All experiments on statistical systems, and in physics at large, are performed at finite times, and general conditions also 
exist under which statistical systems tend to equilibrate to MB energy distributions. This is illustrated,
for instance, in the Boltzmann approach by the H-theorem valid under the assumption of short range interatomic 
interactions. This result is robust with respect to the form of interaction \cite{Khinchin}, which can only affect details of the thermalization, 
most importantly the timescale of the transient. Conversely, there are also results, for specific systems as finite ensembles of harmonic 
oscillators, in which we expect lack of equilibration \cite{Smith}. The most interesting situation consists then on determining energy distributions at finite time. 
Restrictive conditions have been determined for maintaining MB states at all times during an equilibration process 
between two baths \cite{Andersen}. In our previous papers, we have discussed the dynamics of two baths, with distinct species A and B 
consisting of particles of mass $m_A$ and $m_B$ confined in harmonic potentials with trapping angular frequencies $\omega_A$ and $\omega_B$. 
We assume no intra-species interactions at all times. Instead, there are inter-species interactions of the form shown in Eq.~\ref{Hint},  resulting in the total Hamiltonian

\begin{figure*}[t]
\includegraphics[width=0.49\textwidth, clip=true]{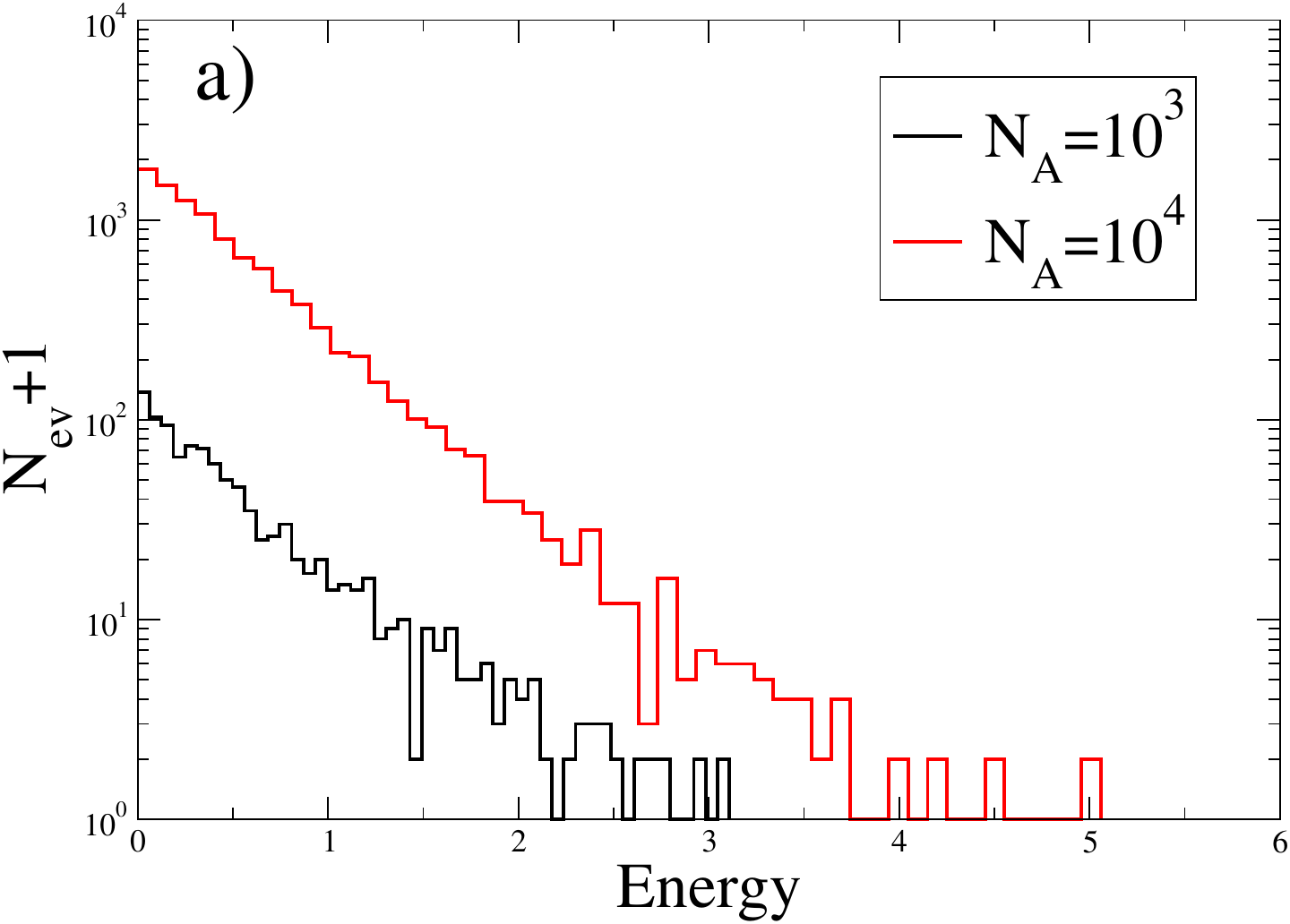}
\includegraphics[width=0.49\textwidth, clip=true]{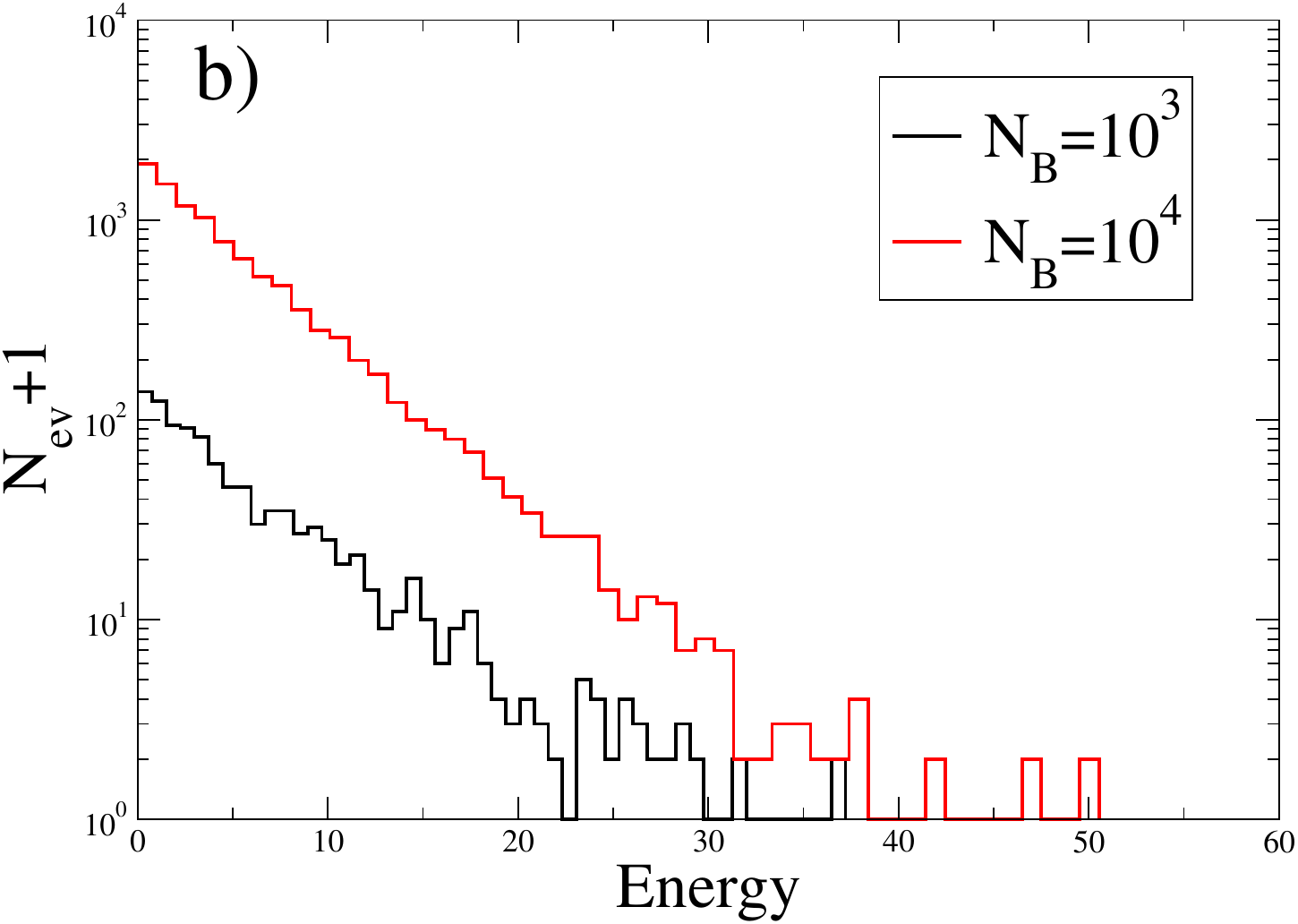}
\caption{Initial energy distributions of the two baths for the populations of $10^3 \times 10^3$ and $10^4 \times 10^4$ particles, generated from MB statistics with inverse temperatures of $\beta_A=2.0$ (left plot) and $\beta=0.2$ (right plot), organized into a number of bins $N_{\mathrm{bins}}=50$ of  equal size between the minimum and the maximum energies.
In the low temperature case of the left panel the best fit of the generated distributions results in values of
 $\beta_A=2.187, N_A= 64.962$ ($\tilde{\chi}^2=0.875$) for the case of $N_A=10^3$ particles, and  
 $\beta_A=1.979, N_A=1126.74$ ($\tilde{\chi}^2=0.966$) for the case of $N_A=10^4$ particles.
In the high temperature case of the right panel the best fit of the generated distribution results in values of 
$\beta_B=0.202, N_B=782.669$ ($\tilde{\chi}^2=2.664)$) for the case of $N_B=10^3$ particles, and 
$\beta_B=0.207, N_B=11129.496$ ($\tilde{\chi}^2=0.945$) for the case of $N_B=10^4$ particles.
If instead we use $N_{\mathrm{bins}}=100$}  for the data of $10^4$ particles, we obtain 
$\beta_A=1.9718, N_A=535.67$ ($\tilde{\chi}^2=1.088$),  
$\beta_B=0.2067, N_B=5274.61$ ($\tilde{\chi}^2=0.800$).
\label{Fig1}
\end{figure*}
\begin{eqnarray}
H&=& \sum_{m=1}^{N_A} H_A(Q_m,P_m)+ \sum_{n=1}^{N_B} H_B(q_n,p_n)+ \nonumber \\
& & \sum_{m=1}^{N_A} \sum_{n=1}^{N_B} H_{\mathrm{int}}(Q_m,q_n),\
\label{Hamilton1}
\end{eqnarray}
where
\begin{equation}
H_A(Q_m,P_m)=\frac{P_m^2}{2m_A}+\frac{1}{2} m_A \omega_A^2 Q_m^2,
\label{Hamilton2}   
\end{equation}
\begin{equation}
H_B(q_n,p_n)= \frac{p_n^2}{2m_B}+\frac{1}{2} m_B \omega_B^2 q_n^2,
\label{Hamilton3}
\end{equation}
and $(Q_m,P_m)$, $(q_n,p_n)$ are the positions and momenta of each particle of the two species $A$ and $B$, respectively. 
We assume the simplest case of one-dimensional systems, although the extension to full three dimensional systems, as 
discussed in \cite{JauOnoSun1,JauOnoSun2}, is possible. This model  recovers the Boltzmann approximation of weak and local interactions 
in the $\gamma, \lambda \to 0$ limit. In the case of weak interactions and large $\lambda$ 
the systems progressively morphs into a set of  linearly interacting harmonic oscillators plus small 
anharmonicities, allowing a direct connection to the so-called Caldeira-Leggett model broadly used for open quantum systems \cite{Magalinskii,Ullersma1,Ullersma2,Ullersma3,Ullersma4,Caldeira1,Caldeira2,Caldeira3}.
\begin{figure*}[t]
\includegraphics[width=0.32\textwidth, clip=true]{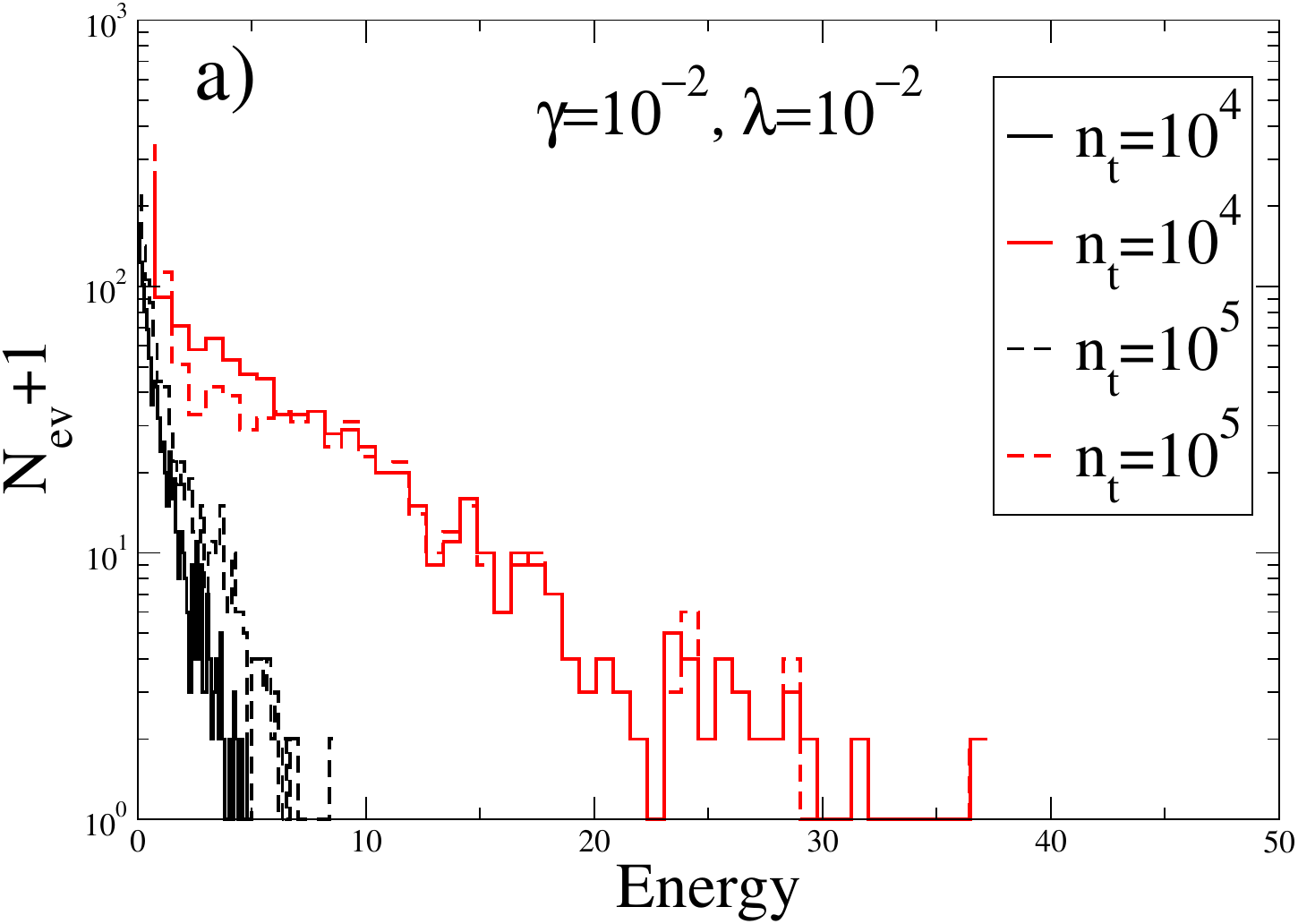}
\includegraphics[width=0.32\textwidth, clip=true]{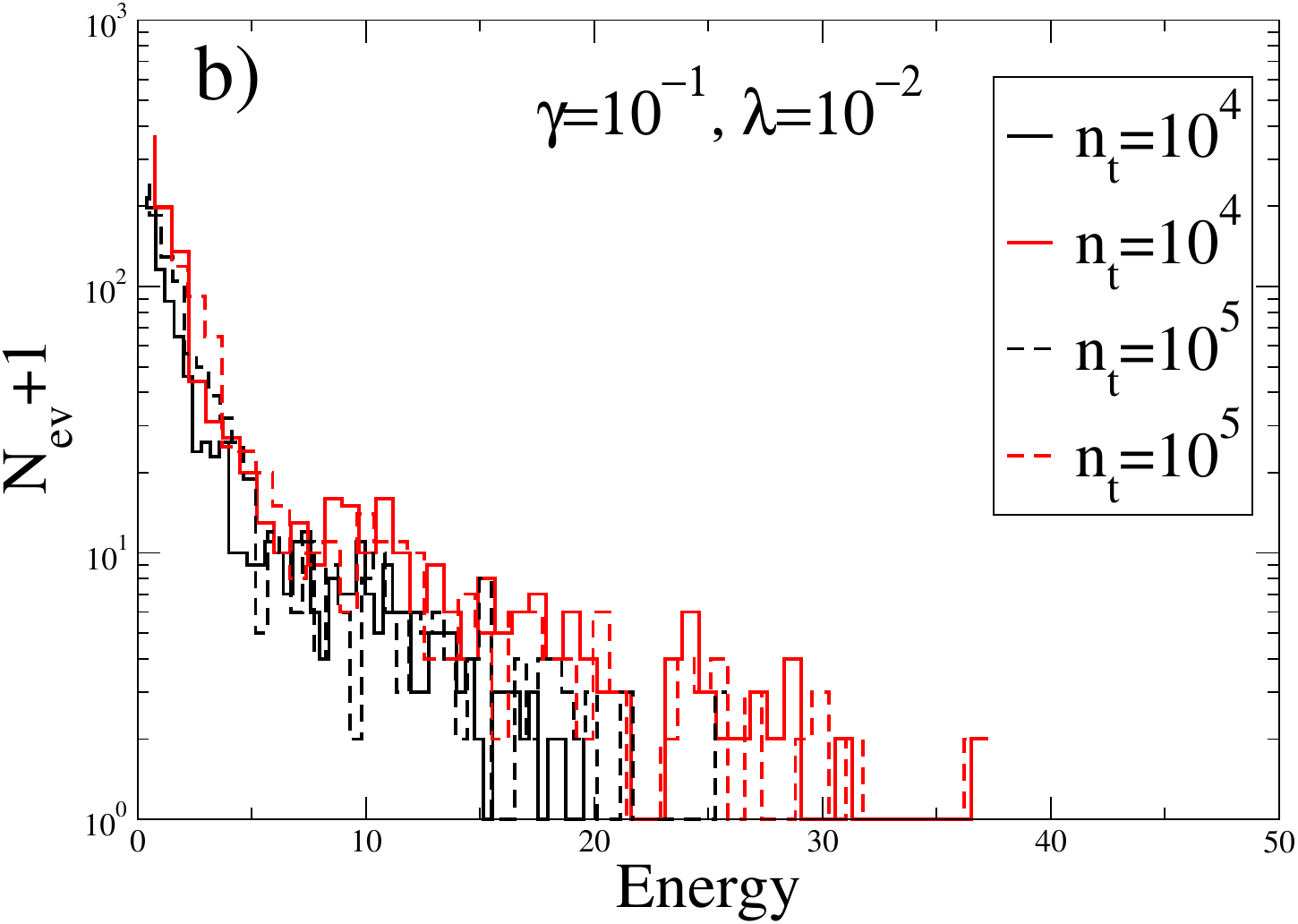}
\includegraphics[width=0.32\textwidth, clip=true]{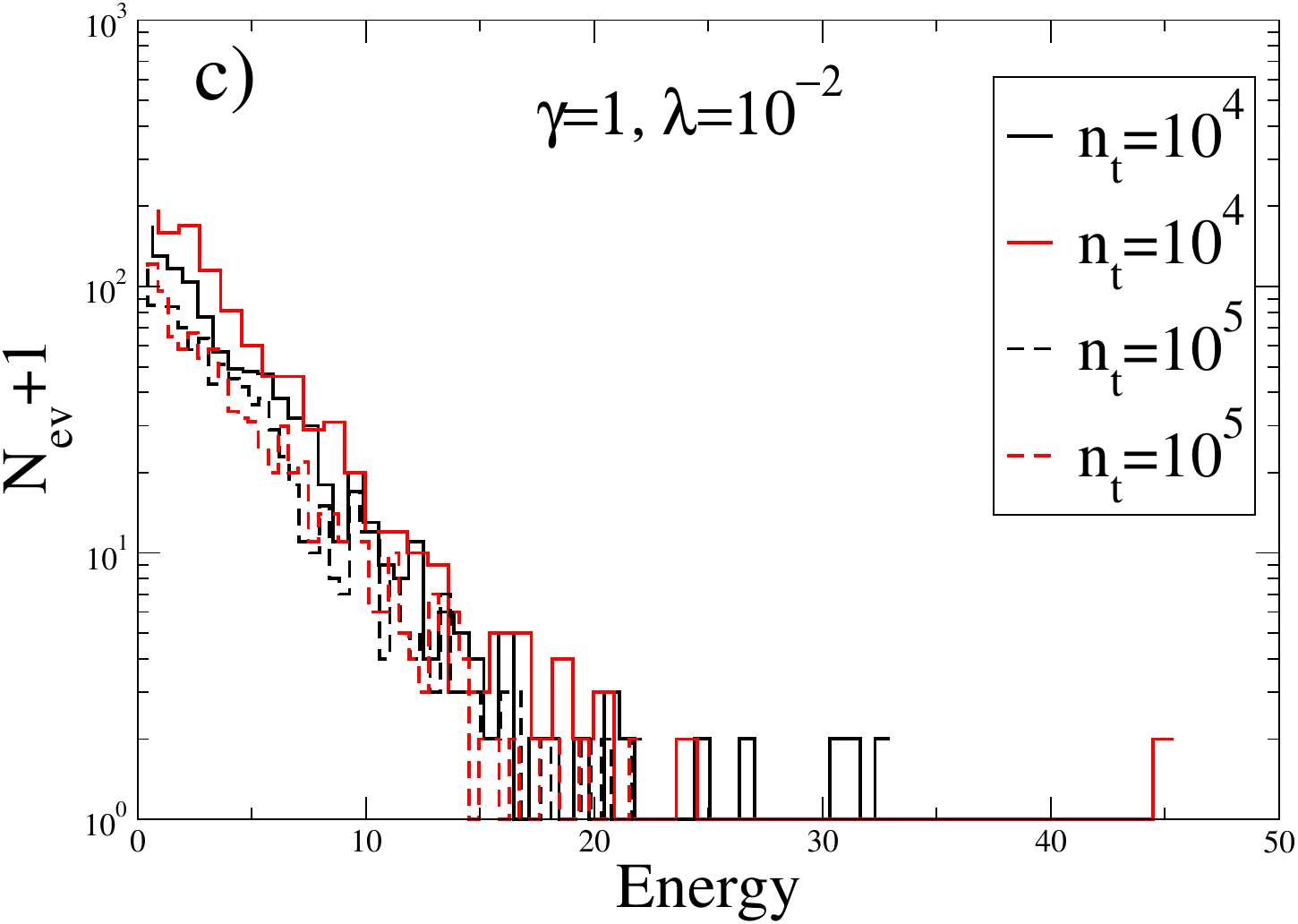}
\includegraphics[width=0.32\textwidth, clip=true]{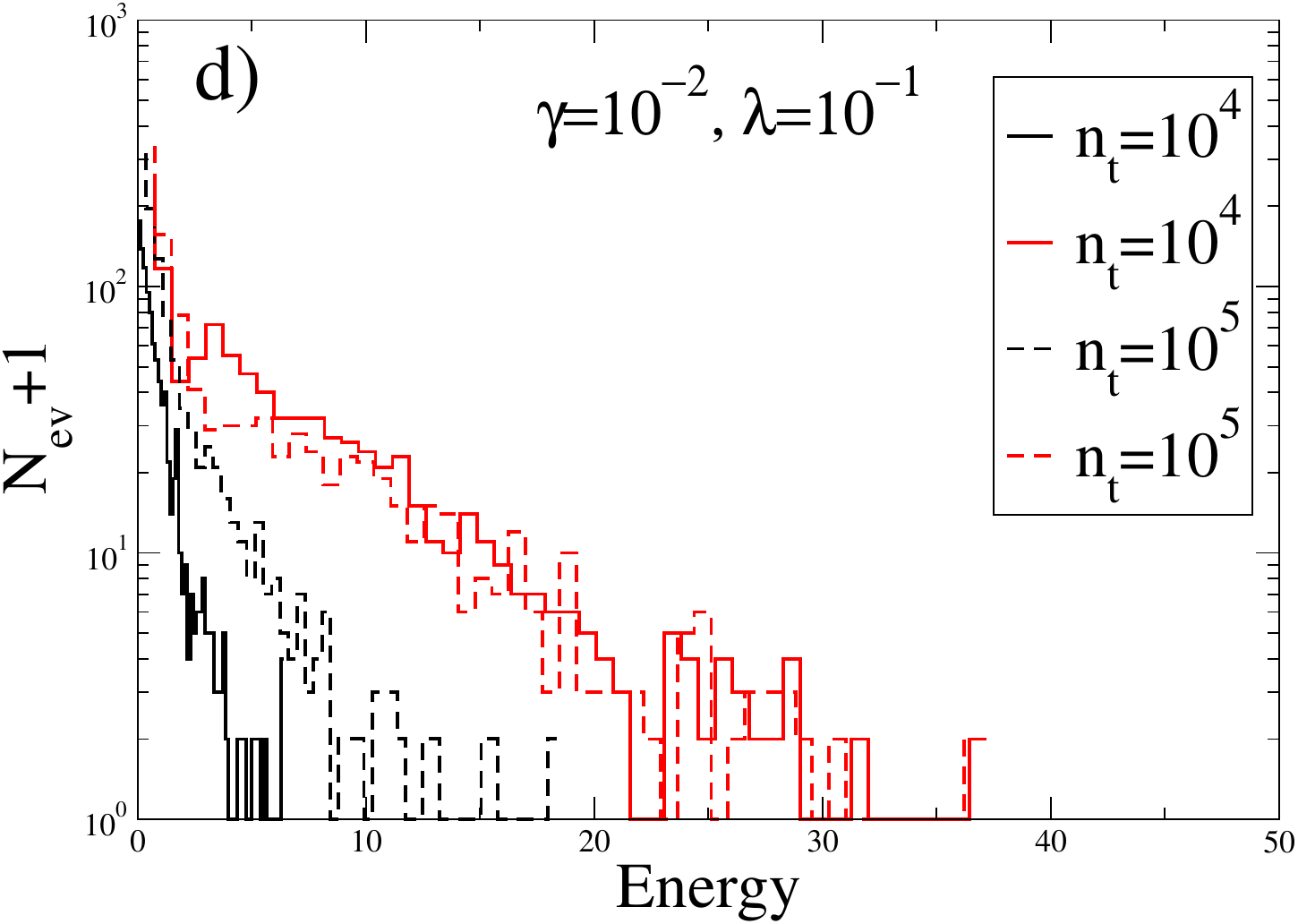}
\includegraphics[width=0.32\textwidth, clip=true]{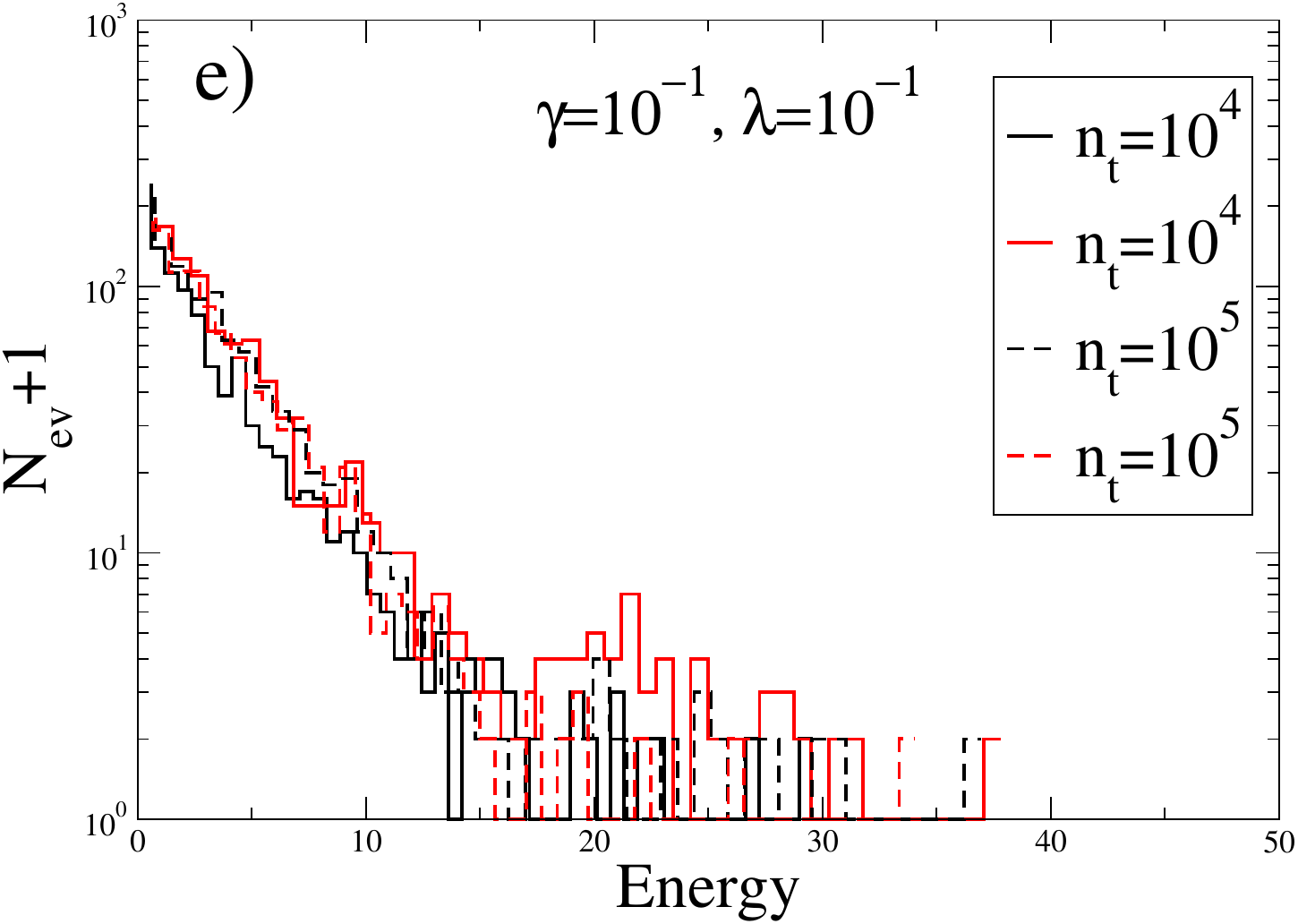}
\includegraphics[width=0.32\textwidth, clip=true]{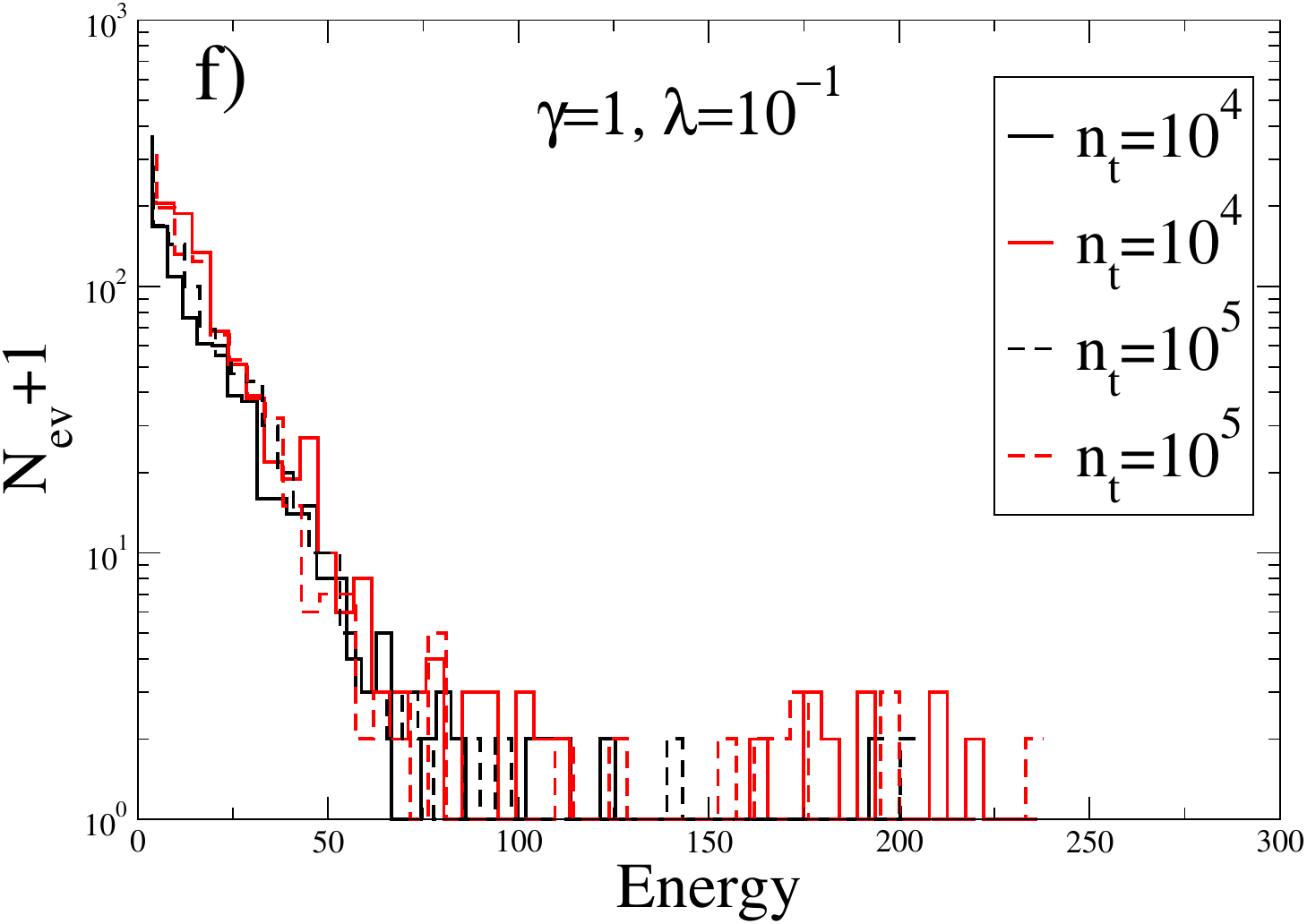}
\includegraphics[width=0.32\textwidth, clip=true]{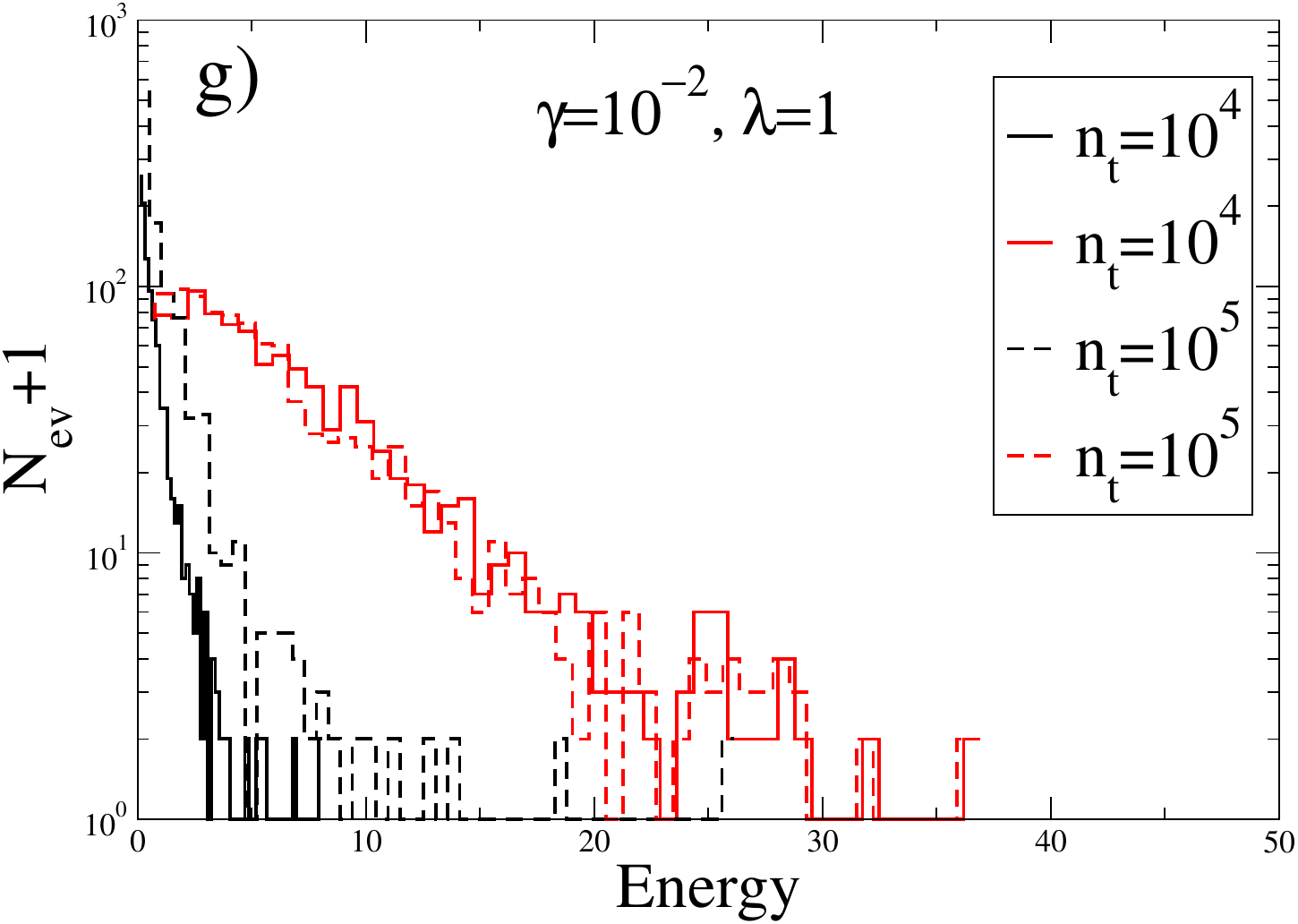}
\includegraphics[width=0.32\textwidth, clip=true]{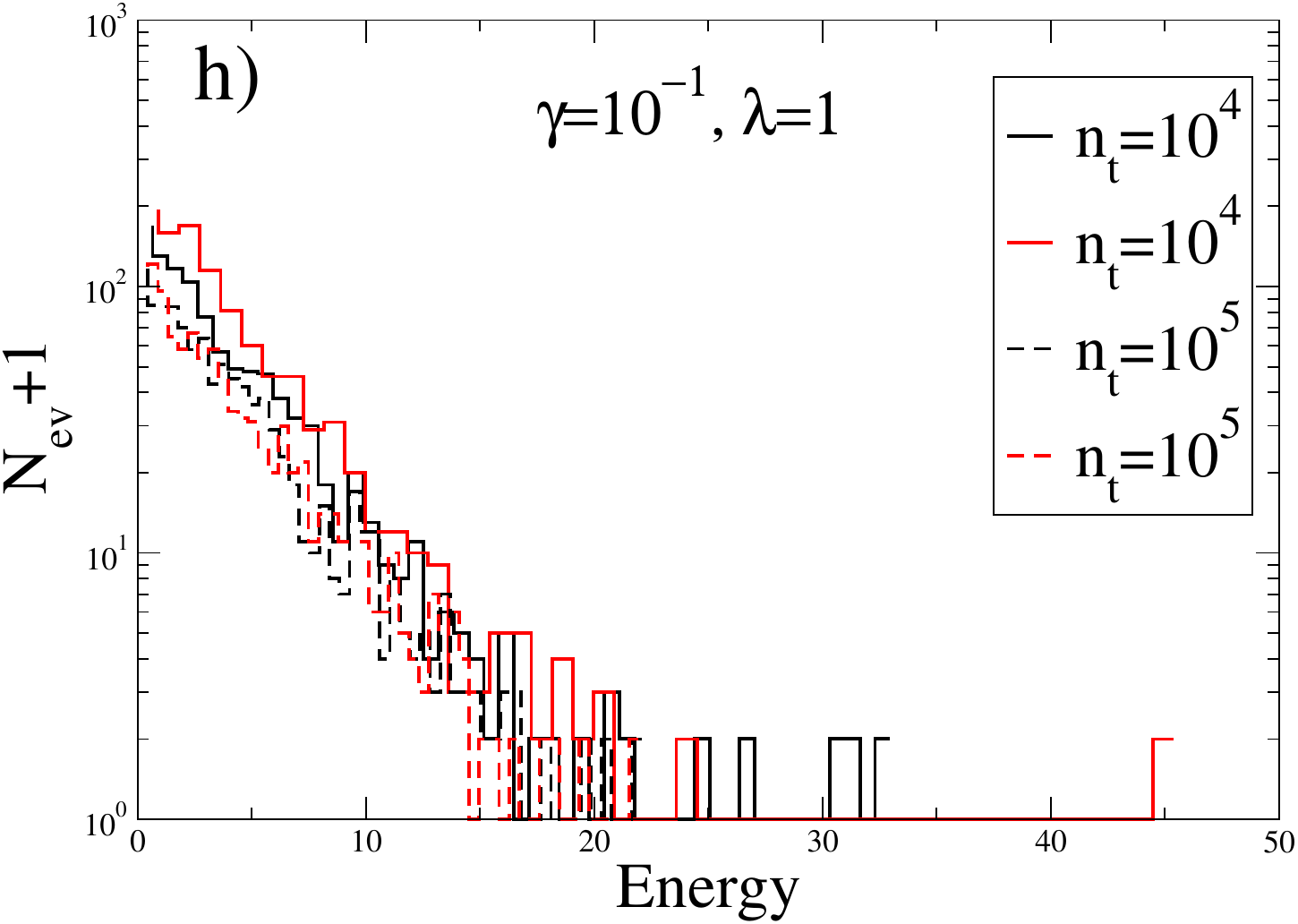}
\includegraphics[width=0.32\textwidth, clip=true]{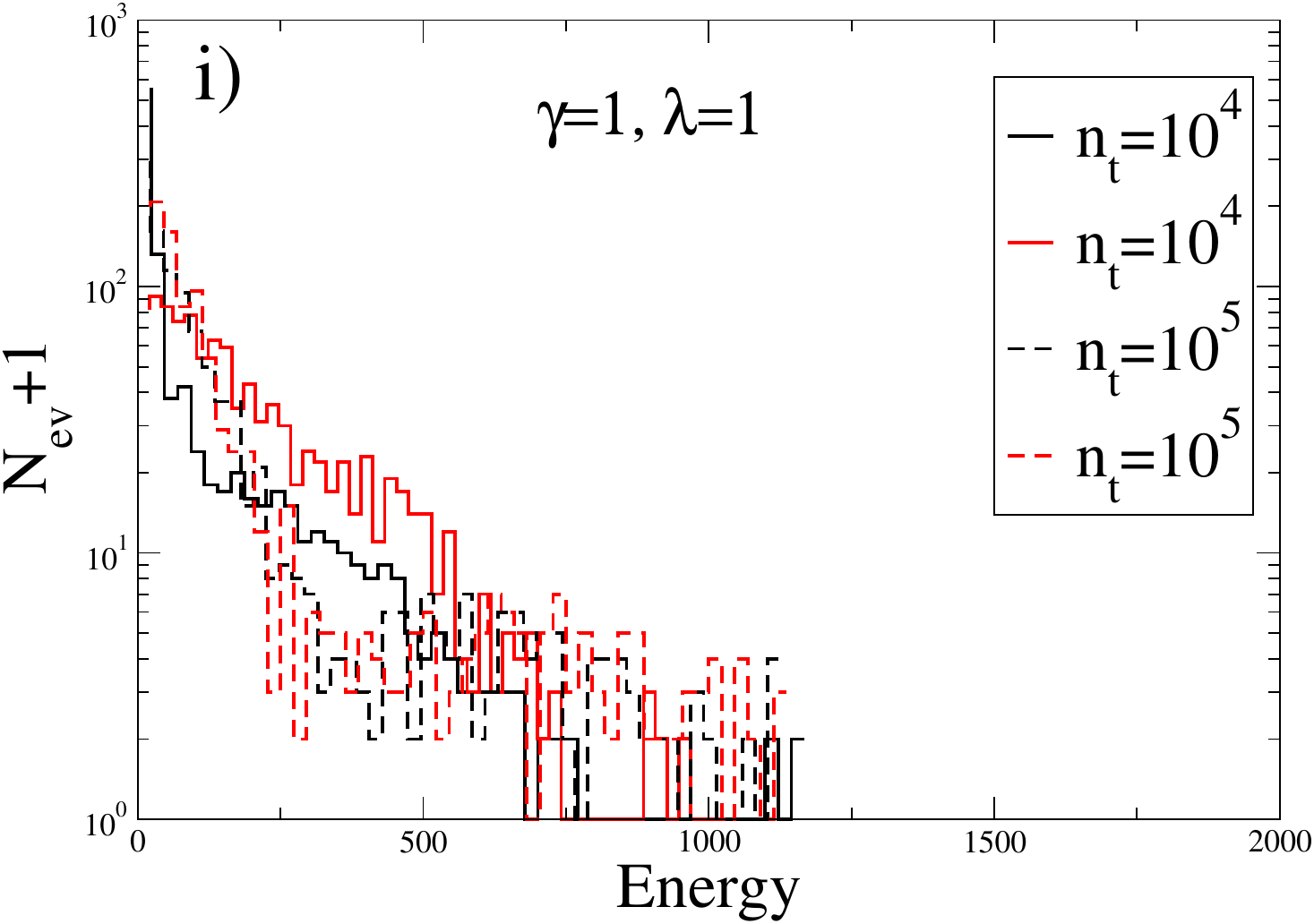}
\caption{Energy probability distributions ($N_{\mathrm{bins}}=50$) evaluated after $n_t=10^4$ and $n_t=10^5$ time steps, for values of $\gamma=0.01, 0.1, 1$ 
and $\lambda=0.01, 0.1, 1$, in the standard situation of equal mass and frequency for the two baths made of $10^3$ particles each. 
The time step
considered is $0.01$ for this and all subsequent plots, and the angular frequencies of the harmonic oscillators are $\omega_A=\omega_B=1$ in all plots apart from the variable frequency ratio case of Fig. \ref{Fig6}.
The vertical axes show the number of particles found with energy between two defined values plus one particle, to allow 
representation in logarithmic scale of the bins with no particles.}
\label{Fig2}
\end{figure*}
The classical equations of motion resulting from $H$ are then numerically integrated, using a variable step predictor-corrector method where
the total system energy is conserved to machine precision. The particle energies in each sub-system are drawn from independent MB distributions of different temperatures.
The harmonic trapping potential for each species then allows for particle positions and momenta to be specified as initial conditions. The sampling time-step, indicated
in the figure caption, is based on the shorter harmonic period of the two species.

This study is numerically challenging for various reasons. With regard to the coupling strength $\gamma$, for small couplings the 
thermalization timescales increase to impractical values, while for large $\gamma$ the differential equation 
solver requires long times for the targeted numerical precision due to the increased stiffness. 
Furthermore, small interaction ranges make interspecies encounters rare, while in the opposite limit the number of interactions becomes large. 
All this is compounded by the statistical errors, particularly prominent in the high energy tail, due to the small number of events.
We have therefore opted for a mixed strategy in which low-statistics, long time simulations have been complemented 
by relatively higher statistics simulations at shorter times. We identify two reasons for possible deviations from the MB 
energy distribution, the strong, non-perturbative regime of the inter-particle interaction at large $\gamma$, and the regime of 
intermediate $\lambda$, even at low $\gamma$, in which the non-extensivity of the interaction energy is prominent. 
\begin{figure*}[t]
\includegraphics[width=0.32\textwidth, clip=true]{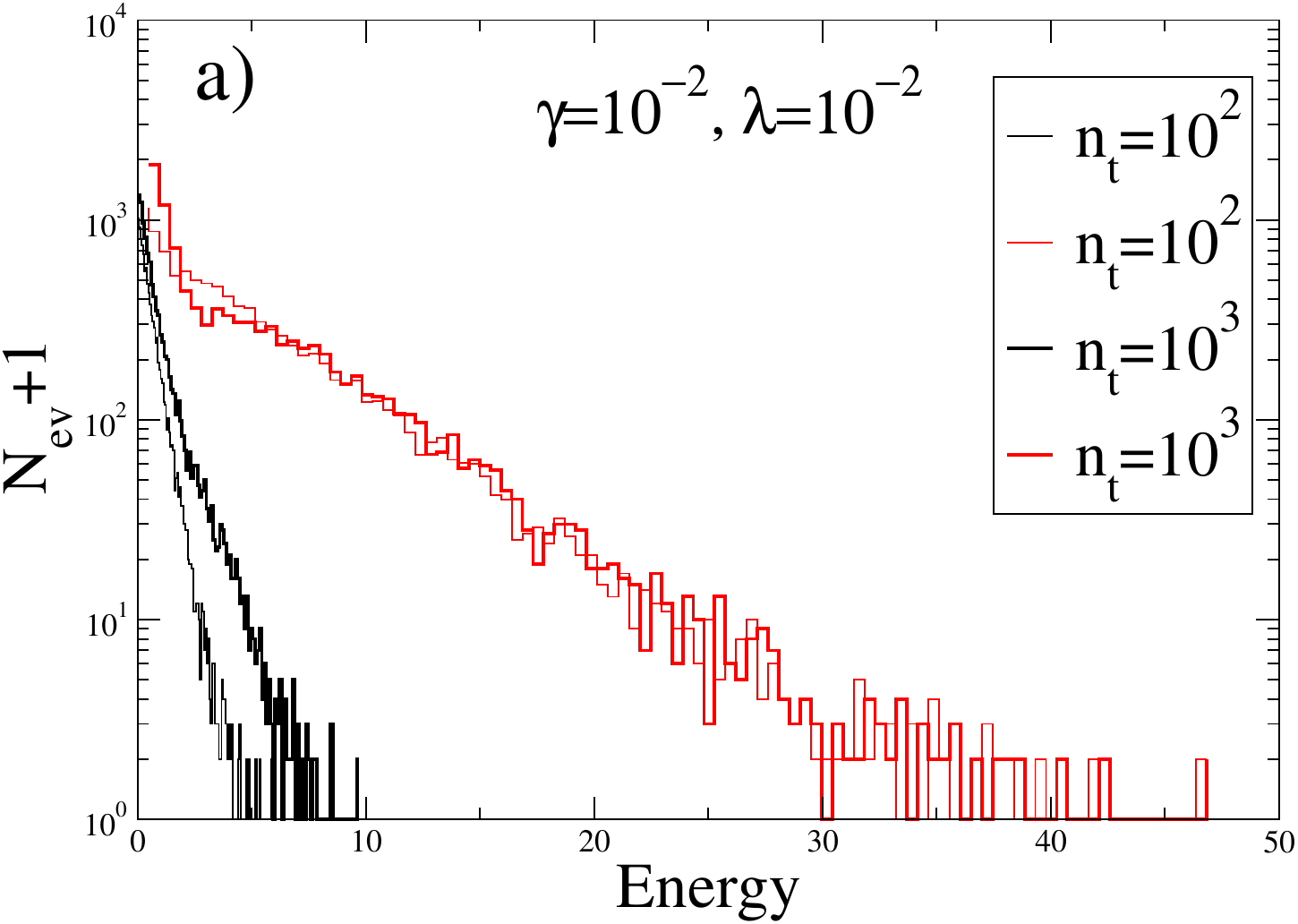}
\includegraphics[width=0.32\textwidth, clip=true]{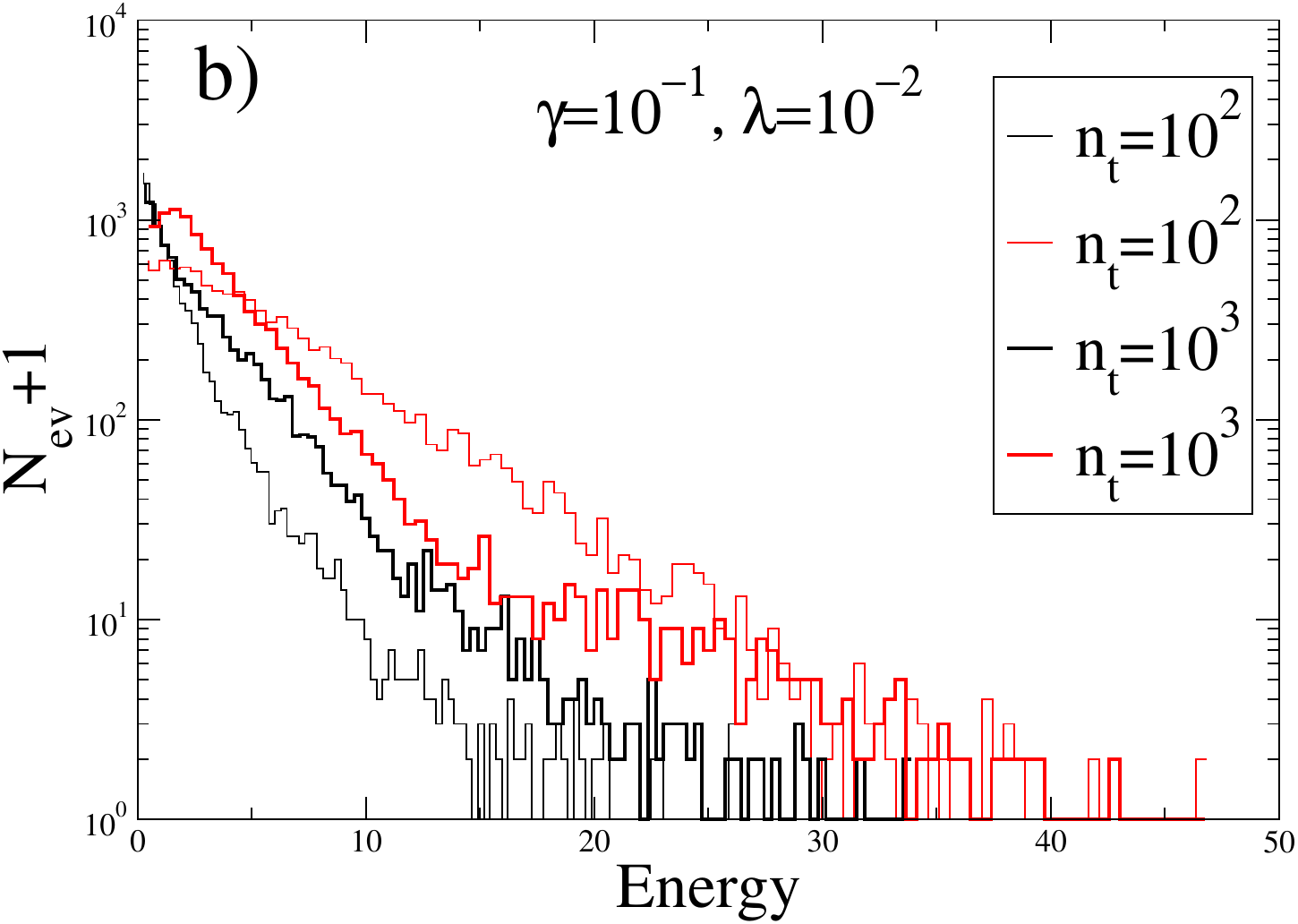}
\includegraphics[width=0.32\textwidth, clip=true]{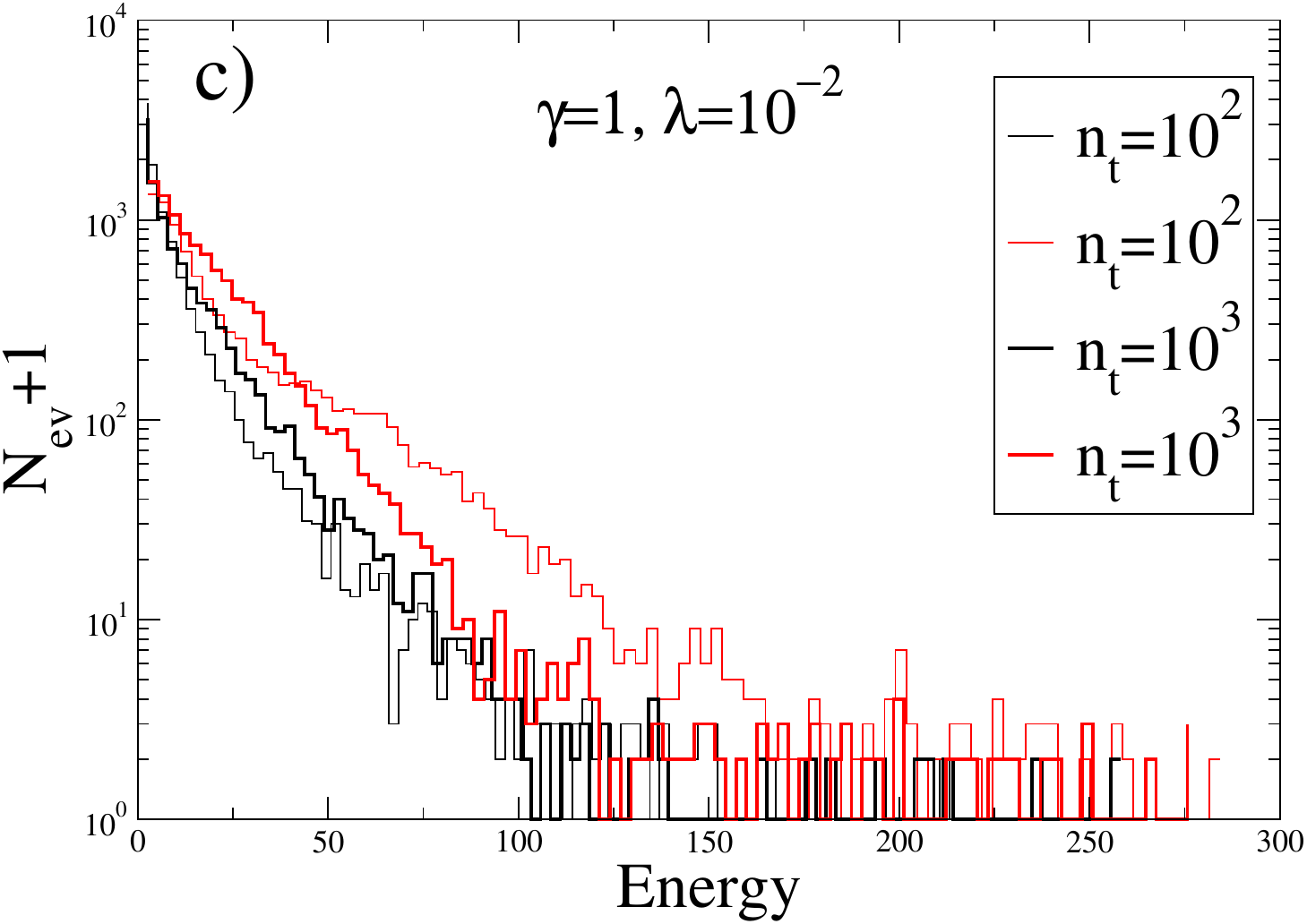}
\includegraphics[width=0.32\textwidth, clip=true]{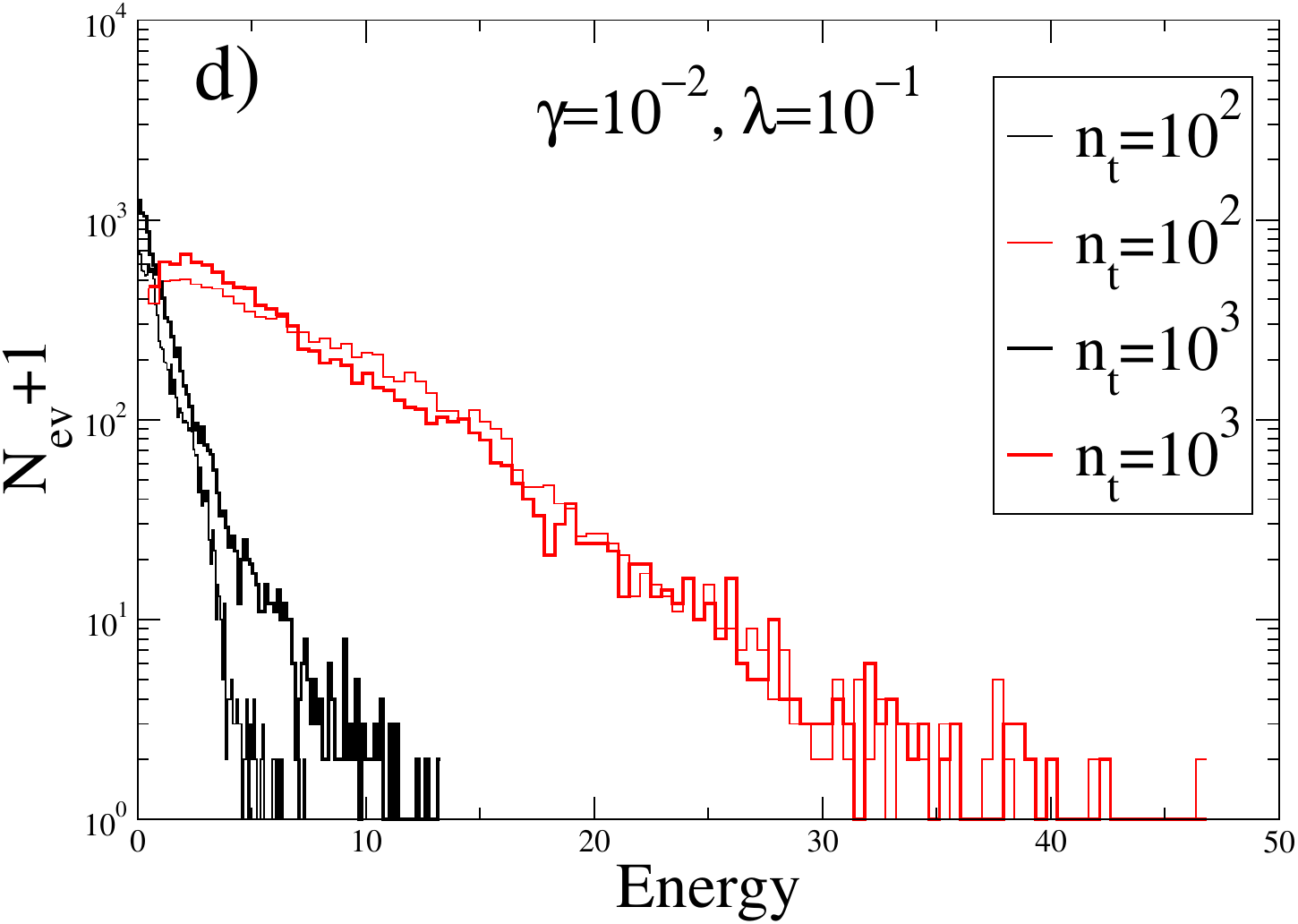}
\includegraphics[width=0.32\textwidth, clip=true]{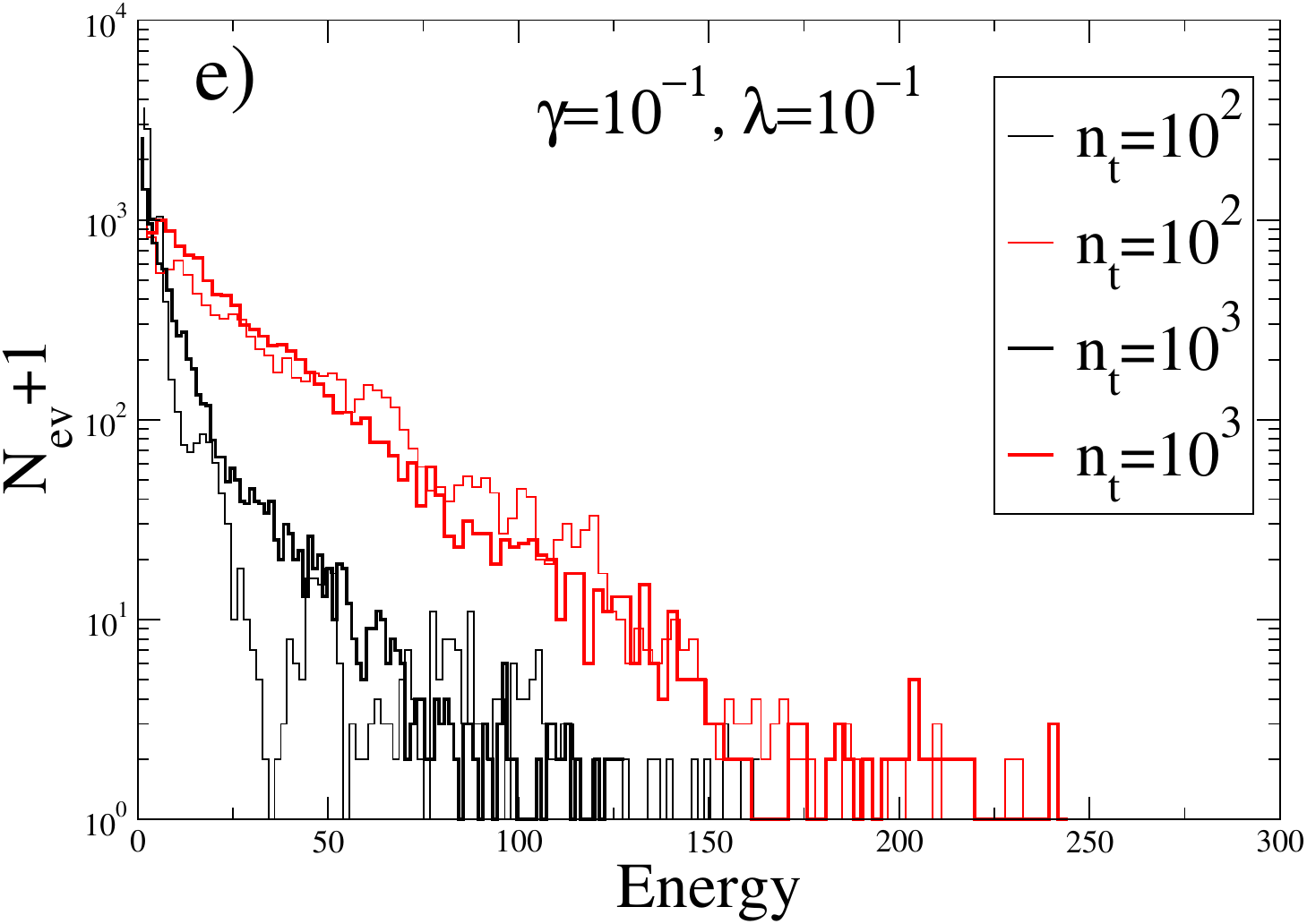}
\includegraphics[width=0.32\textwidth, clip=true]{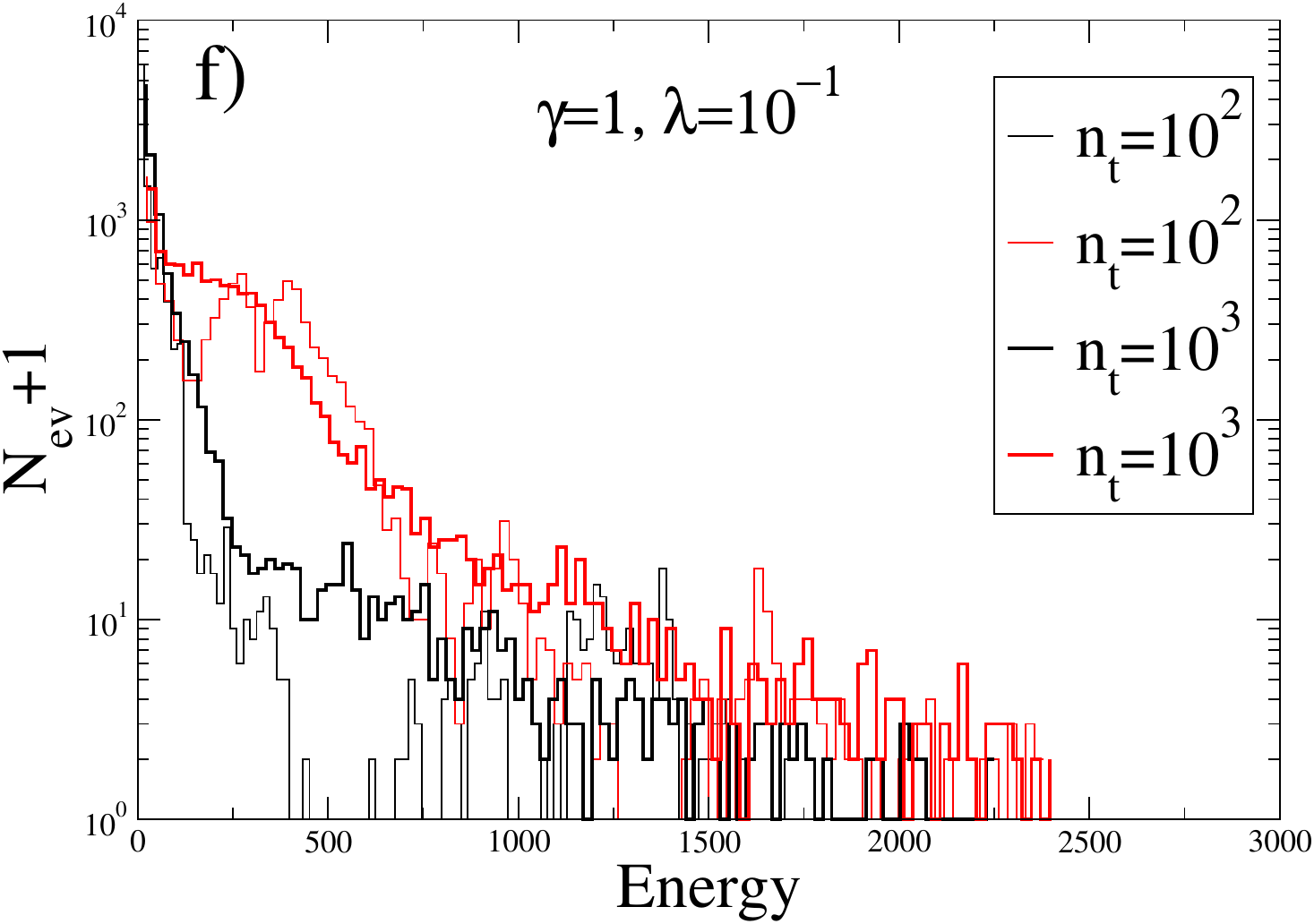}
\includegraphics[width=0.32\textwidth, clip=true]{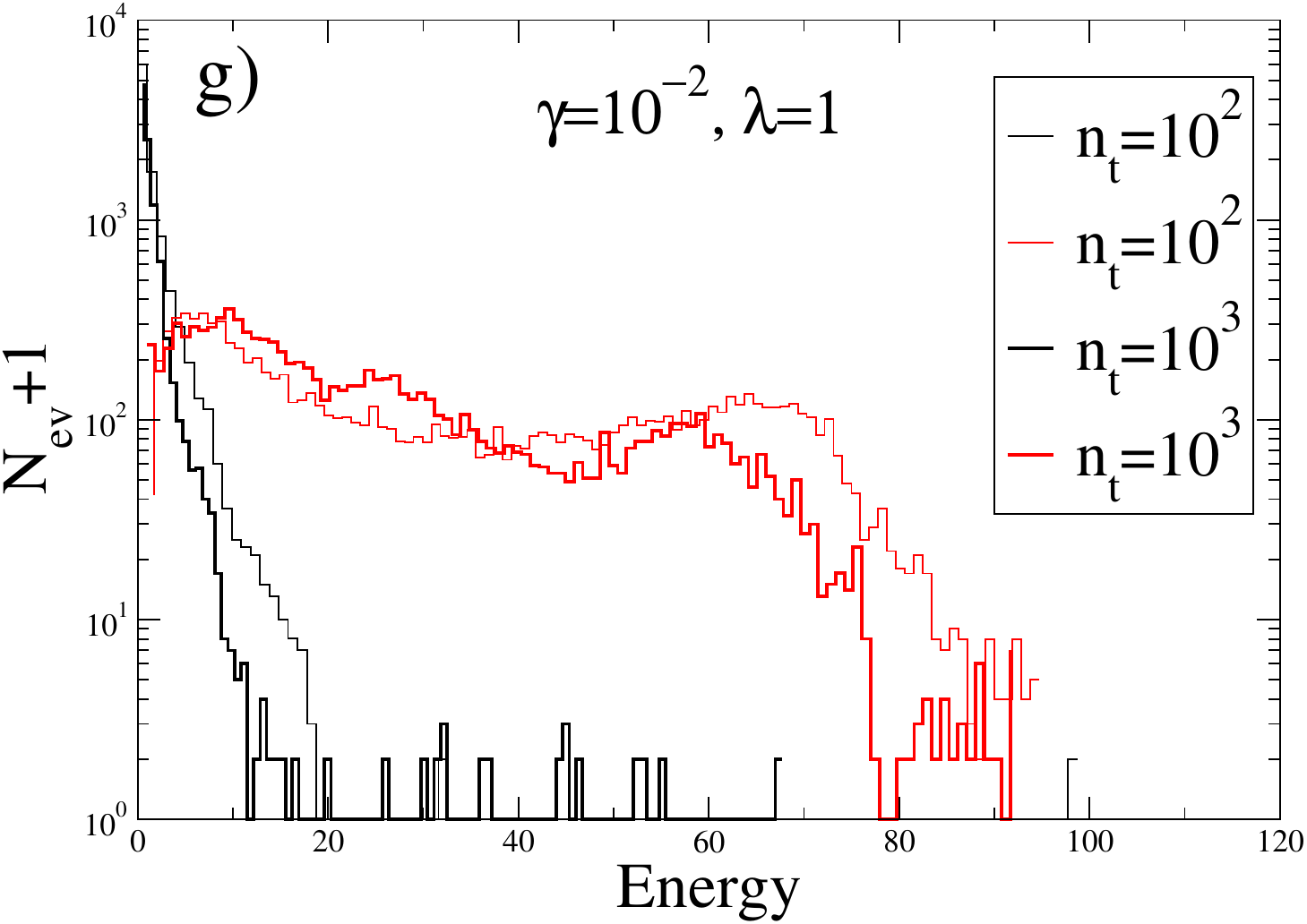}
\includegraphics[width=0.32\textwidth, clip=true]{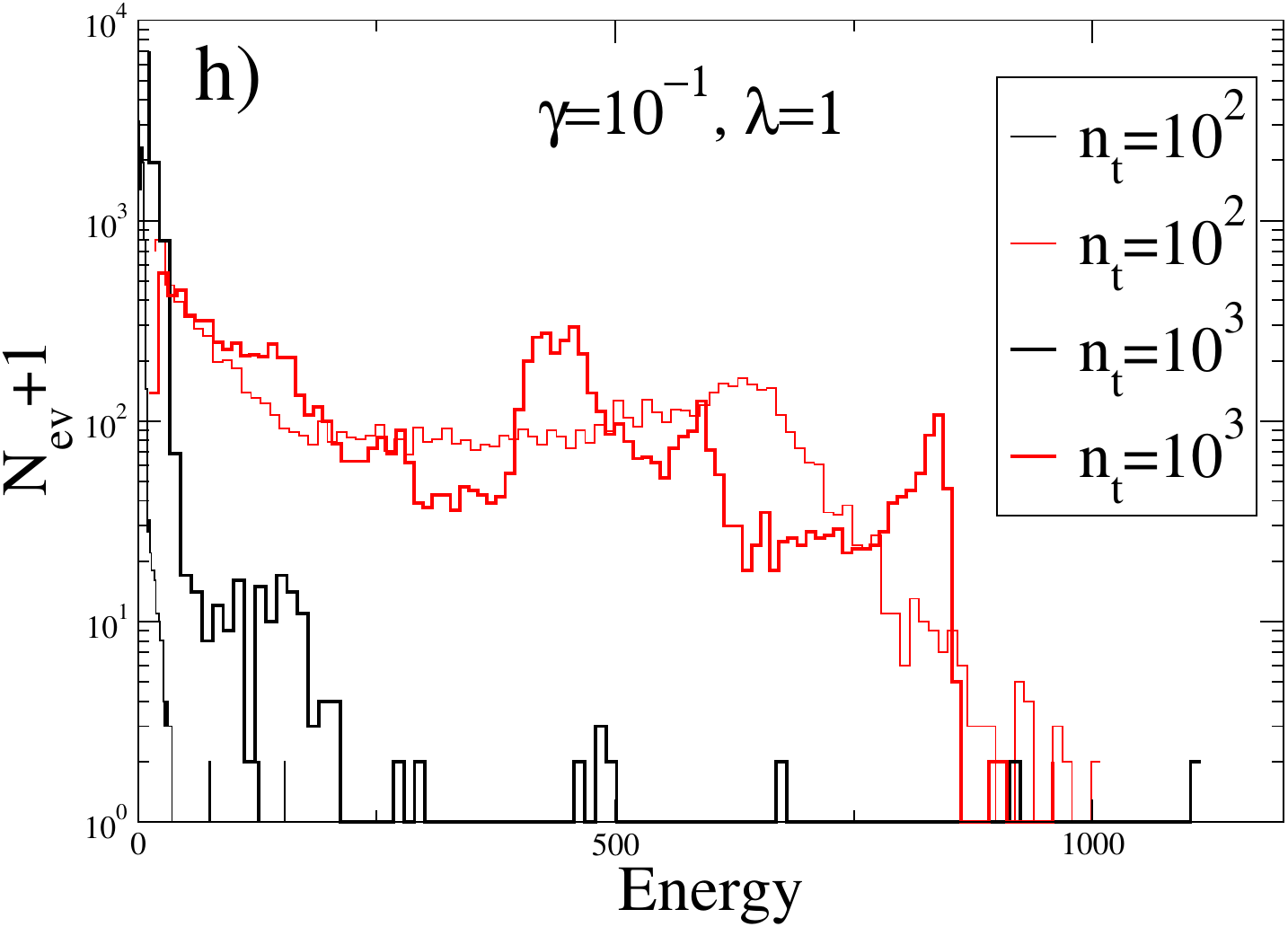}
\includegraphics[width=0.32\textwidth, clip=true]{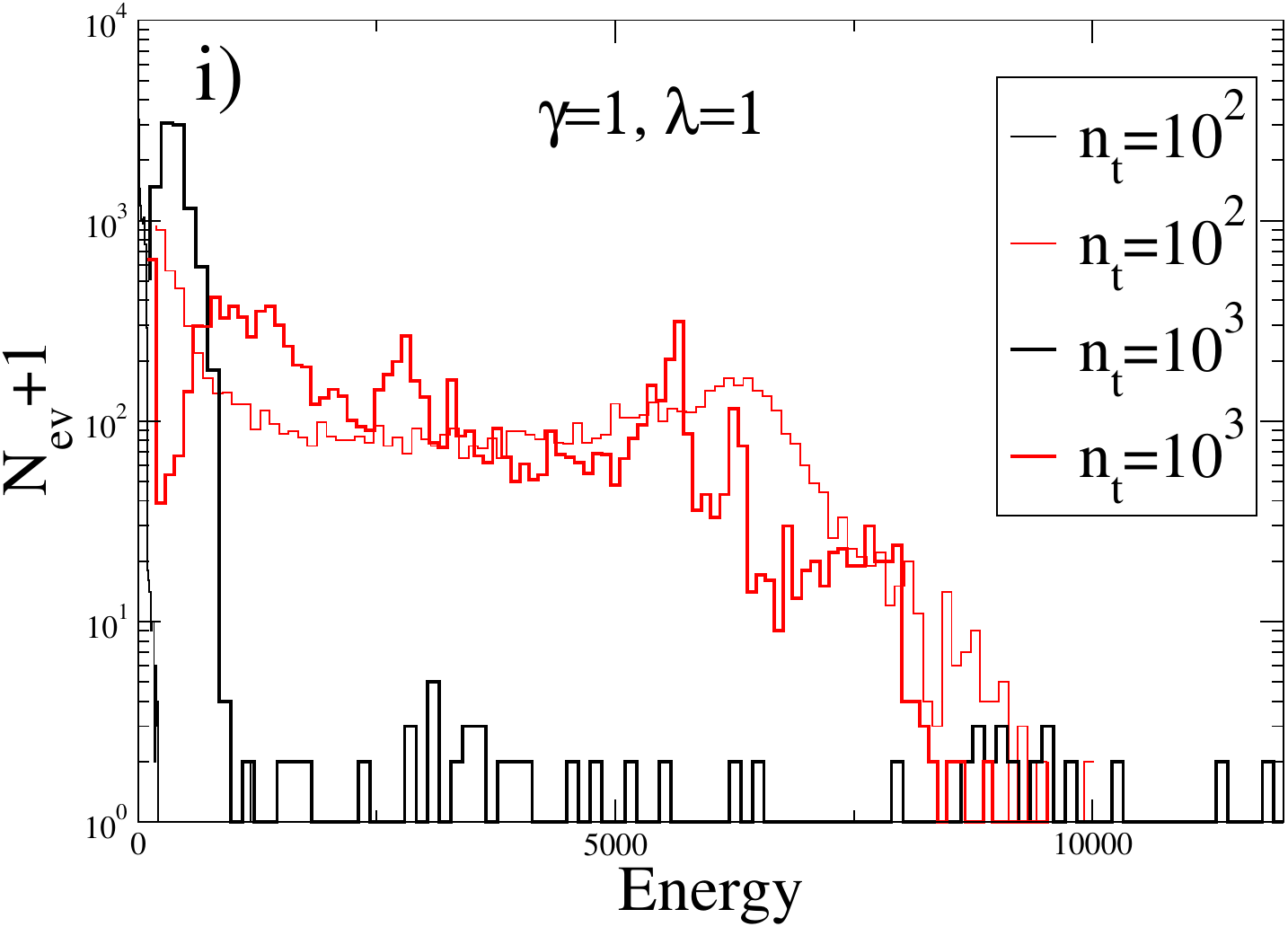}
\caption{Energy probability distributions evaluated after  $n_t=10^2$ and $n_t=10^3$ time steps, for values of $\gamma=0.01, 0.1, 1$ 
and $\lambda=0.01, 0.1, 1$, in the standard situation of equal mass and frequency for the two baths made of $10^4$ particles each, and $N_{\mathrm{bins}}=100$.}
\label{Fig3}
\end{figure*}
As indicated earlier, the initial energies for the particles of each bath are drawn from MB distributions, with probability 
density, normalized to unity and for each species ${i=\mathrm{A,B}}$,  given by
\begin{equation}
p(E_{i})=\beta_i e^{-\beta E_{i}},
\label{Boltzmann}
\end{equation}
where $E_i$ is the total energy of a particle in bath $i$, sum of the kinetic and potential energies,  as made explicit in Eqs. \ref{Hamilton2},\ref{Hamilton3}.
For all the cases shown, the inverse temperatures $\beta_A=0.2$ and $\beta_B=2$ were considered.

The probability density, in a realistic setting with a finite number of particles, needs to be coarse grained. Relative to the total number of 
particles $N$ present in a bath, the number of  `events' $N_{\mathrm{ev}}^{(j)}$ consisting in the number of particles with energy in between $E_j$ and $E_j+\Delta E$ is, in general related to the  probability density $p(E)$ as
\begin{equation}
N_{\mathrm{ev}}^{(j)}= N \int_{E_j}^{E_j+\Delta E} p(E) dE,
\end{equation}
where $\Delta E=(E_{\mathrm{max}}-E_{\mathrm{min}})/N_{\mathrm{bins}}$, with $E_{\mathrm{max}}$ and $E_{\mathrm{min}}$ the maximum and minimum energies measured at a given time, and 
$N_{\mathrm{bins}}$ the chosen number of equal partitions forming the histogram.
 
For ease of visualization, we then plot the quantity $\log(N_{\mathrm{ev}}^{(j)}+1)$ versus the energy of the particle.
The logarithmic choice has the immediate advantage of emphasizing the small number of 
events present at higher energy, with the added unity accounting for the likely possibility that 
some bins may be empty. Moreover, this choice allows for an easier recognition of deviations from the linear dependence of a MB distribution in the lower energy region, where $N_{\mathrm{ev}}^{(j)} \gg 1$. 
 
We first confirmed the sensitivity of this measure for our needs by contrasting results for the number of particles in each thermal bath. We found
that between $10^3$ and $10^4$ balances the
 need for reasonable run times and high statistics in the resulting energy distributions. 
As expected, in the case of relatively small sample of particles, $N_A=N_B=10^3$, there are large relative fluctuations at 
high energies, due to the small number of events. It is worth mentioning that some statistical systems, more specifically 
trapped ultracold atoms, typically work with particle numbers of this order. In order to assess the reliability of the energy distributions 
with respect to the size of their population, in Fig. 1 we contrast the two cases of initial energy distributions with $10^3$ and $10^4$ particles in
each bath, and the related best fit incorporating the MB energy distribution. 
In the case of an MB energy distribution as in Eq. \ref{Boltzmann}, we get, for the particles of the bath at inverse temperature $\beta_i$
\begin{equation}
N_{\mathrm{ev}}^{(j)}=N(e^{-\beta_i E_j}-e^{-\beta_i(E+\Delta E_j)}).
\end{equation}
For the data analysis, we found it more accurate to convert the number of events in each bin into the corresponding probability as 
$\Delta p_j(E)=N_{\mathrm{ev}}^{(j)}/N$ and fit, in the case of a single MB energy distribution, with
\begin{equation}
\Delta p_j^{MB}= e^{-\beta_i E_j}-e^{-\beta_i(E_j+\Delta E)}.
\label{MBprob}
\end{equation} 
Notice that, especially in the case of strong interactions, even if still sufficient to describe an MB energy distribution, 
it is not accurate to expand the last term in Eq.~\ref{MBprob} as $e^{-\beta_i} (1+\beta_i \Delta E)$,
since, in these situations, the validity of the condition $\beta_i \Delta E \ll 1$ is not necessarily verified.

To quantify the accuracy of the fit, our software uses a quantity related to the $\chi^2$ of a fit, and we will adopt a definition of $\tilde{\chi}^2$ as
\begin{equation}
\tilde{\chi}^2=\frac{1}{N_{\mathrm{bins}}}\sum_{j=1}^{N_{\mathrm{bins}}} (\Delta p_j - \Delta p_j^{MB})^2\;,
\end{equation}
which differs from the usual definition of $\chi^2$
\begin{equation}
{\chi}^2=\sum_{j=1}^{N_{\mathrm{bins}}} \frac{(\Delta p_j - \Delta p_j^{MB})^2}{\Delta p_j}, 
\end{equation}
in that we have assumed a variance equal to $\Delta p_j$, as in Poisson statistics, as well as normalized to the number of bins which allows
for an immediate comparison between histograms with different $N_{\mathrm{bins}}$. This choice has the advantage of 
emphasizing the weight of the high energy component, where we generally expect deviations from MB.
Figure 1 shows that we do not see significant differences while increasing the number of particles in each bath. 
Rather, in this case we find values of $\tilde{\chi}^2$ per unit particle larger in the case of the configurations 
with $10^4$ particles in each bath, presumably indicating that the number of bins, $50$ in both cases, should be adjusted
for larger samples. This is an interesting digression and we plan to address the optimization of the statistical binning in a future contribution.

With the setting described above, we performed a variety of simulations, of which a representative sample is shown in Fig.~\ref{Fig2}. 
The choice of $\lambda$ and $\gamma$ values is determined on one hand from the computation time, which, in order to show
thermalization, implies that very  small values of these two parameters should be avoided. Moreover, if $\gamma$ is too large, 
an instability occurs at the center of the trap, and a bistable potential emerges, as discussed in \cite{OnoSun1}, which leads to 
longer thermalization times. 

Deviations from MB are expected, in our model, due to various sources. First, we expect deviations when the coupling 
constant $\gamma$ is large enough to make the interaction energy of the two baths comparable or larger than the 
intrinsic energy (kinetic+potential) of each bath. Second, we expect deviations when considering interaction ranges 
such that multiple interactions of one particle with many in the other bath are not negligible. This occurs approximately when 
$\lambda$ exceeds the average distance between two particles of different baths. Finally, we expect faster thermalization in the 
central region of the trap, where the density of the two species is maximal. Then there will be 
a transient in which the two slopes of the energy distributions approach each other at low energy, while this is suppressed in
the high energy tails. 

These expected qualitative features are confirmed in the plots, which are aligned for constant $\gamma$ along 
columns and constant $\lambda$ along rows. In the situation closer to the Boltzmann assumption of 
small interaction range and interaction strength, panel on the top-left, the thermalization process is quite slow and seems to 
proceed via MB intermediate states. The temperature in the coldest bath appears to have increased, for the same time steps, by increasing the 
interaction range, seen from comparing panel (a) with panels (d) and (g). If the interaction strength is increased instead, thermalization is initially more 
efficient, as seen from  panels (a) and (b), but eventually the large interaction energy leads to deviations from the MB distribution, 
see panel (c). Notice also, in panel (b), that thermalization is nearly complete at low energy, while the high-energy tails of the two baths 
are not yet overlapping. Finally, the effect of the interaction energy is most prominent in the bottom-right panel, (i), notice the scale on the horizontal axis. Therefore the qualitative 
features discussed in the previous paragraphs are confirmed by the simulations. 
\section{Equilibration at short times}
For applications, a detailed study of the thermalization process at short times is important. For instance, this understanding could
be crucial in 
temporarily accelerating the dynamics of fusion processes in confined plasmas,  resulting in more favorable 
conditions to ignite self-sustained burning. We present this specific situation in Fig.~\ref{Fig3}, in which snapshots of the 
energy distribution are taken at $10^2$ and $10^3$ time steps, for different values of 
$\gamma$ and $\lambda$. To reduce the statistical errors, we have decided to boost the number of particles by a 
factor of 10 in each bath, and in Fig.~\ref{Fig1} (left plot) we show the initial energy distribution of the baths for the case with $10^4$ particles.
\begin{table*}[t]
\begin{center}
\begin{tabular}{| c | c | c | c | c | c || c | c | c | c | c | c | c | c |}
\hline
$\lambda$ & $\gamma$ & $\beta_A$ & $\beta_B$ & $\tilde{\chi}_A^2$ & $\tilde{\chi}_B^2$ & $\beta_A^{(1)}$ & $\beta_A^{(2)}$ & $\beta_B^{(1)}$ & $\beta_B^{(2)}$ 
& $\tilde{\chi}_A^2$ & $\tilde{\chi}_B^2$ & $p_1^{(A)}$ & $p_1^{(B)}$ \\
\hline
0.01   & 0.01 & 1.760 & 0.588 & $6.9 \times 10^{-6}$  & $8.3 \times 10^{-5}$  & 1.00    & 0.98   &  1.001 & 0.18 & $6.8 \times 10^{-5}$  & $3.4 \times 10^{-5}$ & 0.904 & 0.559  \\
0.01   & 0.1   & 0.526 & 0.312 & $1.4 \times 10^{-5}$  & $3.1 \times 10^{-5}$  & 0.528  & 4.99   &  0.996 & 0.31 & $1.6 \times 10^{-5}$  & $3.2 \times 10^{-5}$ & 0.969 & 0.996  \\
0.01   & 1      & 0.170 & 0.073 & $1.2 \times 10^{-4}$  & $2.7 \times 10^{-5}$  & 0.186  & 0.177 &  0.073 & 0.07 & $1.3 \times 10^{-4}$  & $1.6 \times 10^{-5}$ & 0.628 & 0.235  \\
0.1     & 0.01 & 1.195 & 0.166 & $7.8 \times 10^{-6}$  & $1.6 \times 10^{-5}$  & 1.188  & 1.229 &  0.164 & 0.17 & $7.8 \times 10^{-6}$  & $1.6 \times 10^{-5}$ & 0.947 & 0.894  \\
0.1     & 0.1   & 0.284 & 0.048 & $5.6 \times 10^{-5}$  & $8.5 \times 10^{-6}$  & 0.292  & 0.899 &  0.048 & 0.07 &  $6.1 \times 10^{-5}$ & $8.5 \times 10^{-6}$ & 0.927 & 0.993  \\
0.1     & 1      & 0.025 & 0.005 & $6.2 \times 10^{-4}$  & $4.0 \times 10^{-5}$  & 0.025  & 0.083 &  0.0048 & 0.039 &  $6.2 \times 10^{-4}$ & $3.5 \times 10^{-5}$  & 0.990 & 0.936  \\
1        & 0.01 & 0.841 & 0.036 & $7.3 \times 10^{-4}$  & $9.3 \times 10^{-3}$  & 1.005  & 0.018 & 0.035  &  0.0355 &  $7.7 \times 10^{-4}$ & $1.1 \times 10^{-5}$ & 0.991 & 0.956 \\
1        & 0.1   & 0.061 & 0.0037 & $2.2 \times 10^{-3}$& $5.1 \times 10^{-5}$  & 0.063  & 0.056 & 0.0085 & 0.0019  & $2.2 \times 10^{-3}$  & $4.8 \times 10^{-5}$ & 0.980 & 0.381 \\
\hline
\end{tabular}
\caption{Analysis of the deviations from the MB energy distribution. The table presents, for all the cases of $\lambda$ and $\gamma$ in Fig. 3 and at $10^3$ time steps, the values of the inverse temperatures of the two baths for the best fit with a single MB, with the related $\tilde{\chi}^2$, and the values of the inverse temperatures of the two baths 
for a best fit with a bMB distribution, including the weight $p_1$ of the dominant MB term, and the  $\tilde{\chi}^2$ of the best fit. All the cases start with initial MB energy distributions, nominally with $\beta_A=2.0, \beta_B=0.2$, which correspond to a best fit for a single MB yielding $\beta_A=2.173$ with a $\tilde{\chi}^2=5.8 \times 10^{-6}$, and $\beta_B=0.221$ with a $\tilde{\chi}^2=4.2 \times 10^{-6}$, while fitting with a bMB provides the same inverse temperatures and approximately the same $\tilde{\chi}^2$.}
\end{center}
\label{Table1}
\end{table*}
These data are presented in the same fashion as in Fig.~\ref{Fig2}, with the three rows corresponding to the 
increasing coupling strength $\gamma$ from left to right, and the same interaction range $\lambda$, and the three columns with increasing $\lambda$ 
from top to bottom, and the same $\gamma$. In the left column, there is no noticeable change in going from $\lambda=10^{-2}$ to 
$\lambda=10^{-1}$, as the two distributions, due to the weak coupling between the two baths, do not have time to start thermalizing 
even after $10^3$ time steps. In the case of $\lambda=1$, both energy distributions become broader and with significant 
deviations from the MB one, still far from thermalizing.  The latter phenomenon is confirmed in the middle and right columns,
where the plots in the bottom (h) and (i) are quite similar to the bottom-left (g), apart from the energy scales which are boosted by a factor 5 and 10, respectively. 
This shows that deviations from an MB energy distribution are mainly attributable to the finite interaction range, with the coupling strength 
only playing the role of linearly rescaling the energy. In addition, the top row shows that if the range is kept constant and small, the effect of the 
coupling constant is minor, and the two energy distributions approach each other with MB profiles when $\gamma$ is increased. 
\begin{figure}[b]
\includegraphics[width=0.95\columnwidth, clip=true]{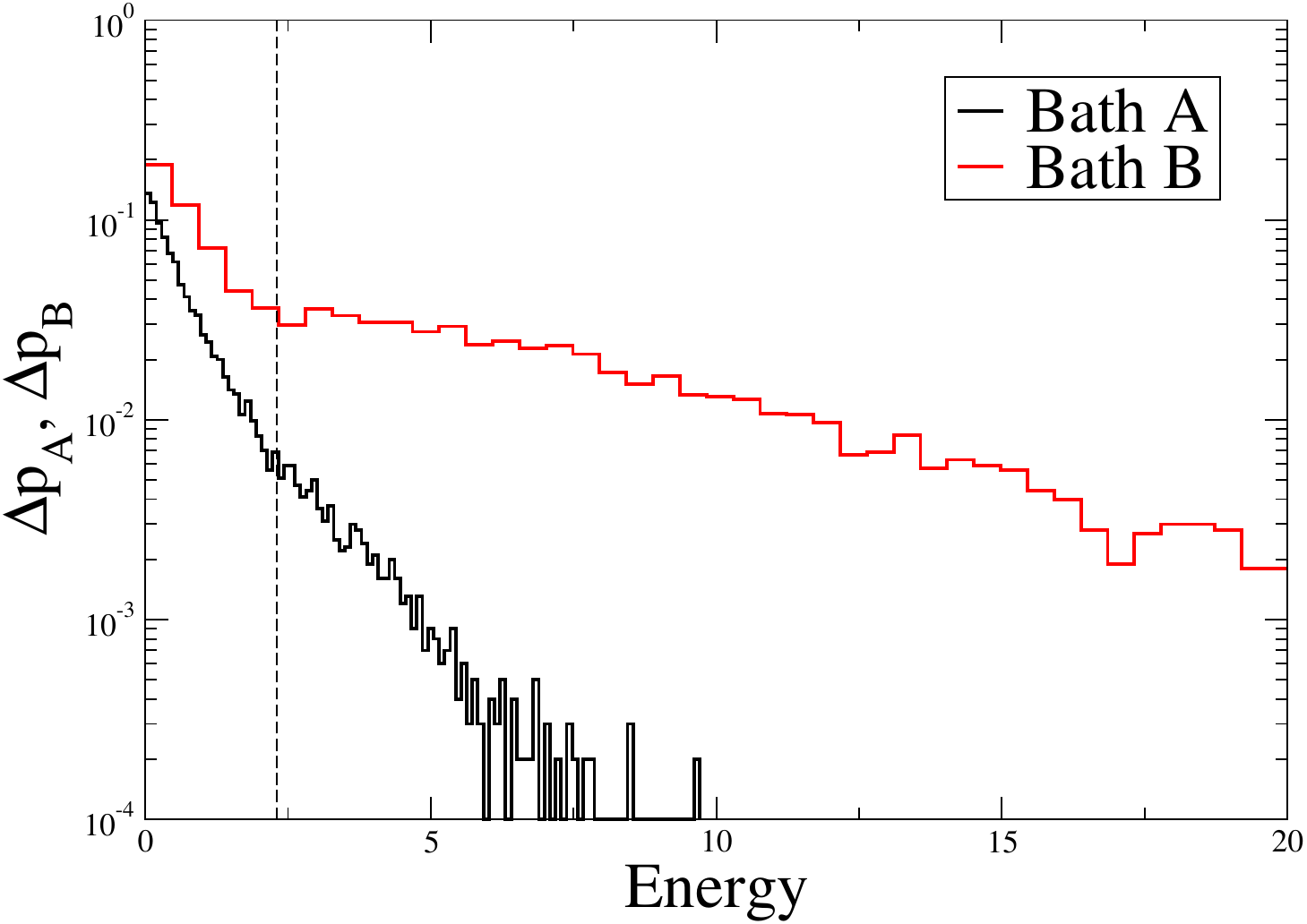}
\caption{Probability distributions of the two baths in the case of $\lambda=10^{-2}, \gamma=10^{-2}$, top row of Table I, 
showing the bimodal feature of bath B due to the strong cooling experienced by its particles in the center of the trap, where the largest 
population of the low temperature species A results in efficient cooling of species B. Vice versa, the lower temperature species A is efficiently heated 
by the hotter species B in the outer region of trapping, corresponding to the higher energy populations.  
The vertical dashed line indicates the approximate energy at which the kinks, corresponding to the sudden change of slopes in the bMB distributions, occur for both baths.}
\label{Fig4}
\end{figure}
In a number of contexts,  the correlation between the presence of  long-range interactions and so-called kappa energy distributions in space physics \cite{Vasyliunas,Treumann,Leubner1,Leubner2,Livadiotis1,Livadiotis2,Livadiotis3}, also known as $q$-distributions in non-extensive statistical mechanics, has been 
shown~\cite{Plastino,Rapisarda,Tsallis,Marin}. These distributions have a hard high-energy tail and therefore their distinctive feature is an excess of events 
at high energy with respect to the corresponding MB distribution. Unfortunately it is not simple to  demonstrate the presence of Kappa distributions with 
numerical simulations involving a limited number of particles.  The most prominent difference from a MB energy distribution 
is expected at high energy, where the statistics of events is rather small. However, it is possible to attempt a comparison between the usual MB and a two-temperature or 
bimodal version (bMB), and  explore its dependence on the $\gamma$ and $\lambda$ parameters. This, apart from being a convenient and simple parameterization of 
deviations from a pure MB energy distribution, is also related to the emergence of kappa distributions because, as discussed in \cite{Hahn}, the latter can be expanded 
as a convolution of MB distributions, of which the simplest is the bimodal case. 
\begin{figure*}[t]
\includegraphics[width=0.32\textwidth, clip=true]{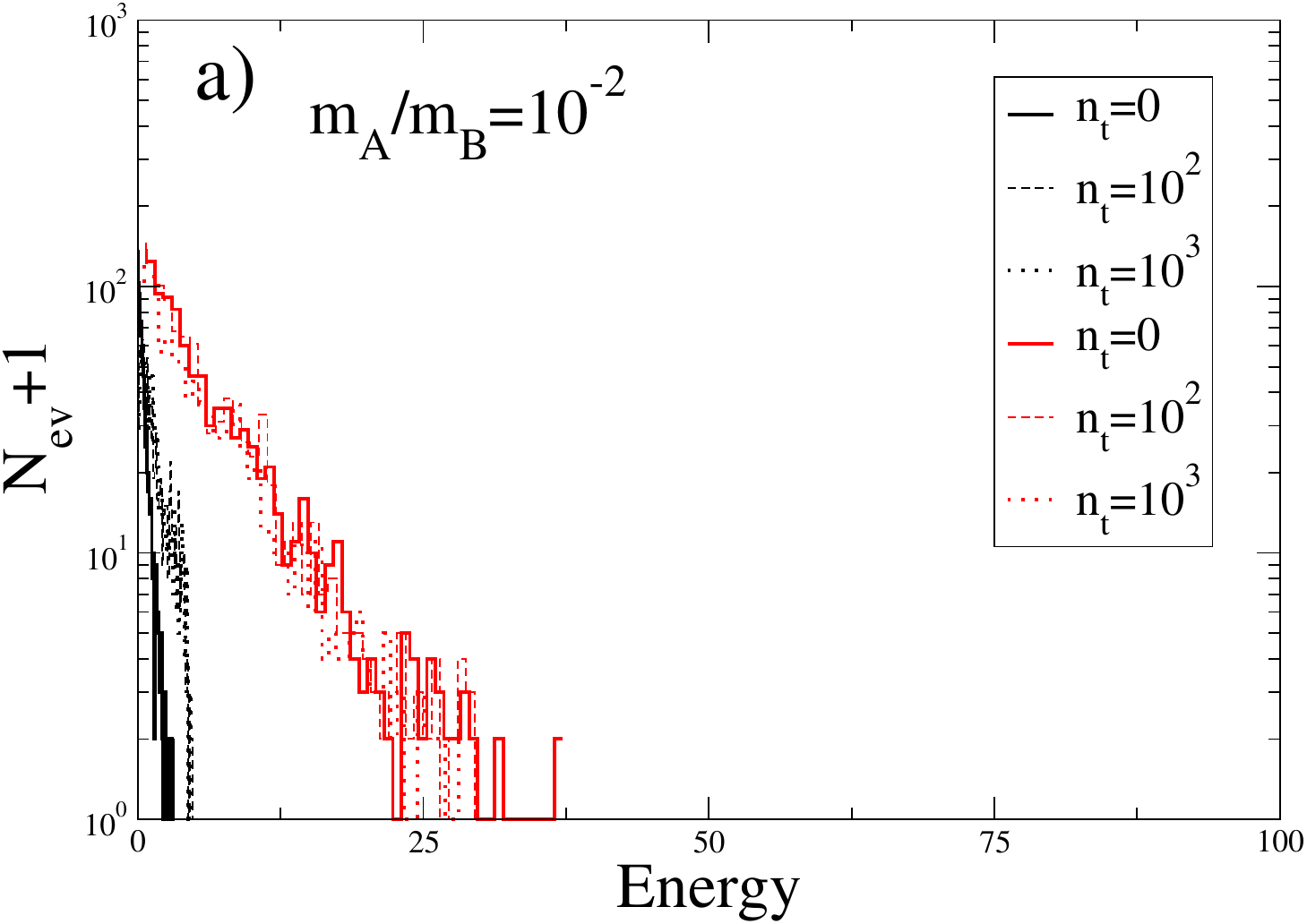}
\includegraphics[width=0.32\textwidth, clip=true]{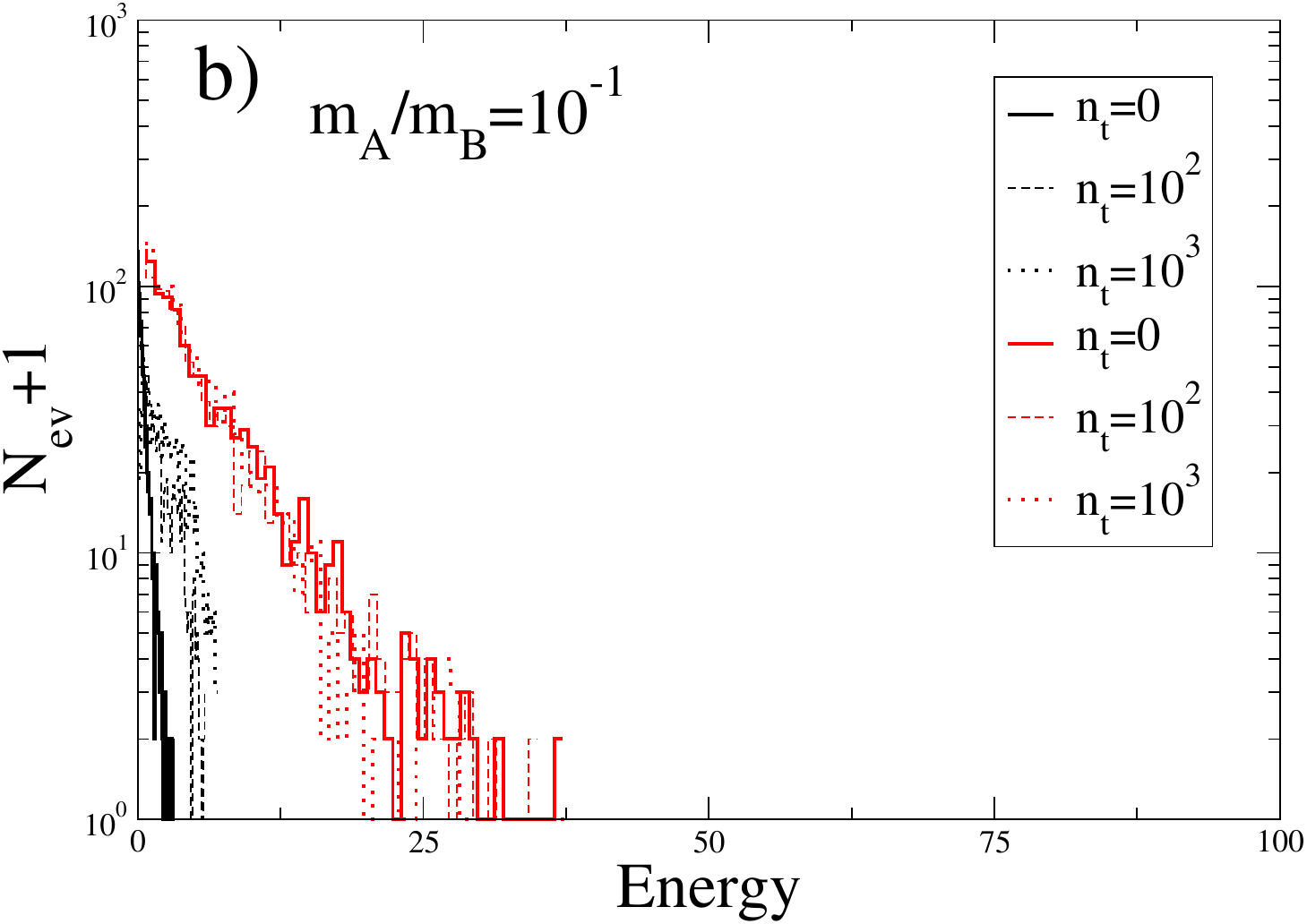}
\includegraphics[width=0.32\textwidth, clip=true]{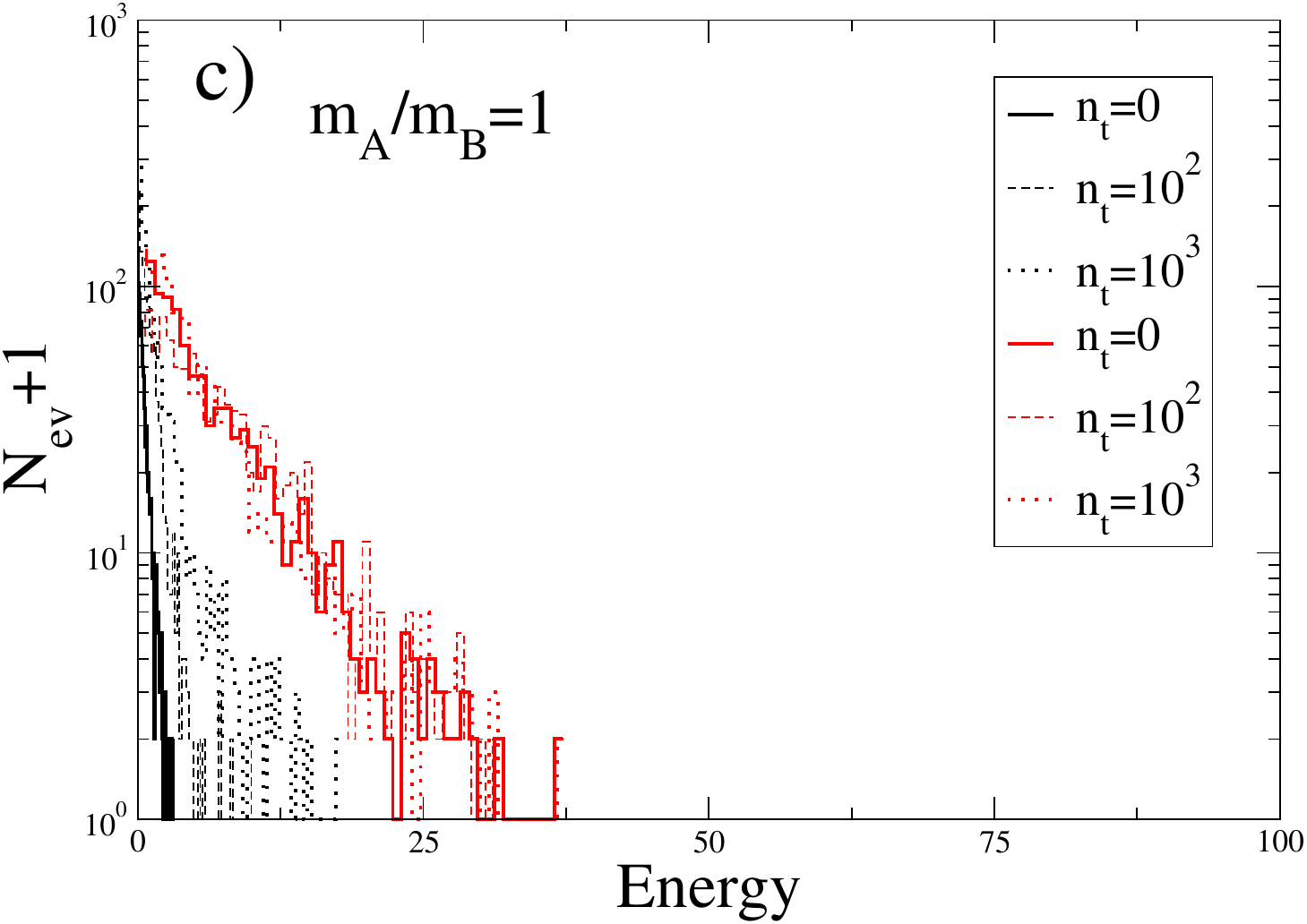}
\includegraphics[width=0.32\textwidth, clip=true]{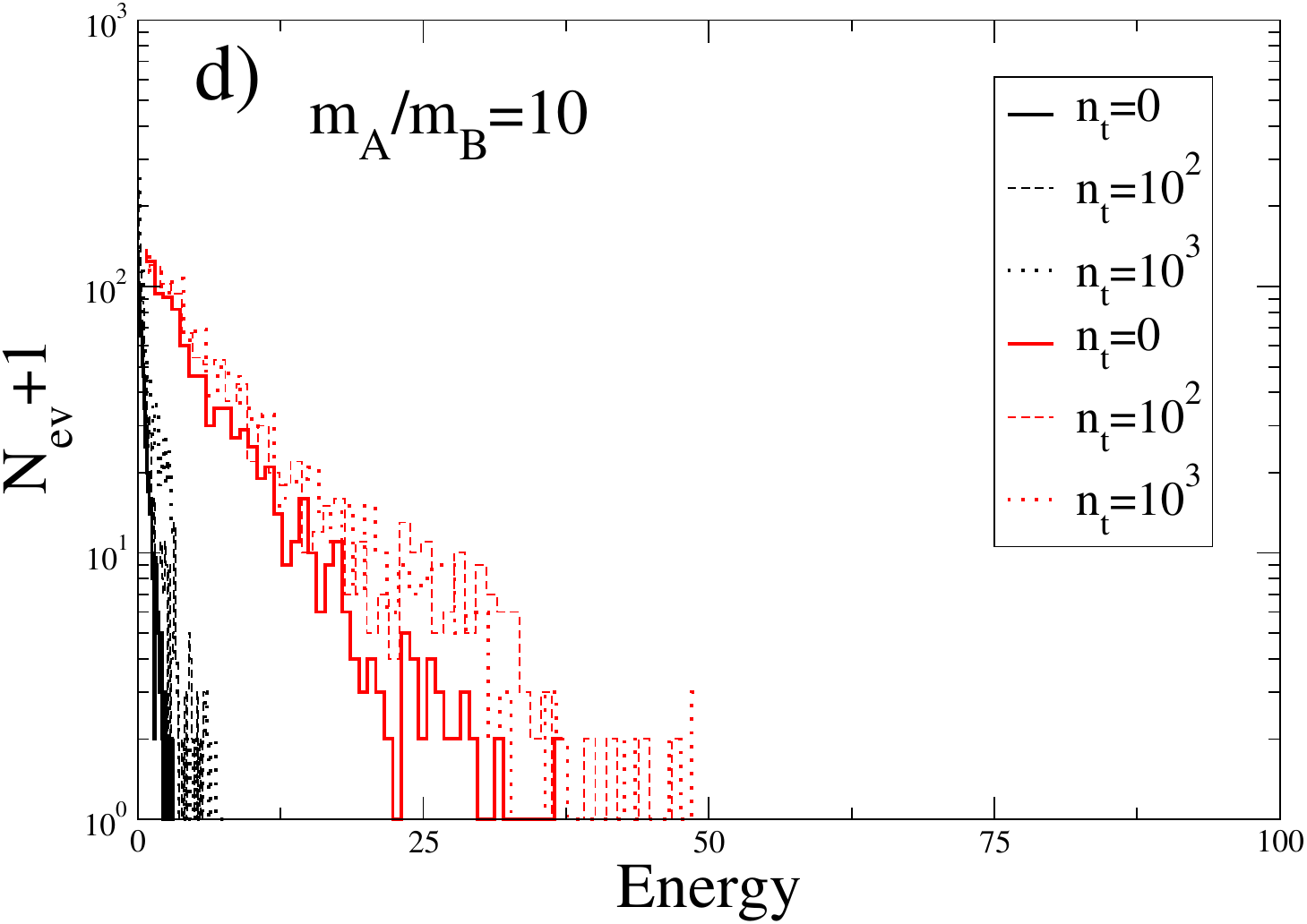}
\includegraphics[width=0.32\textwidth, clip=true]{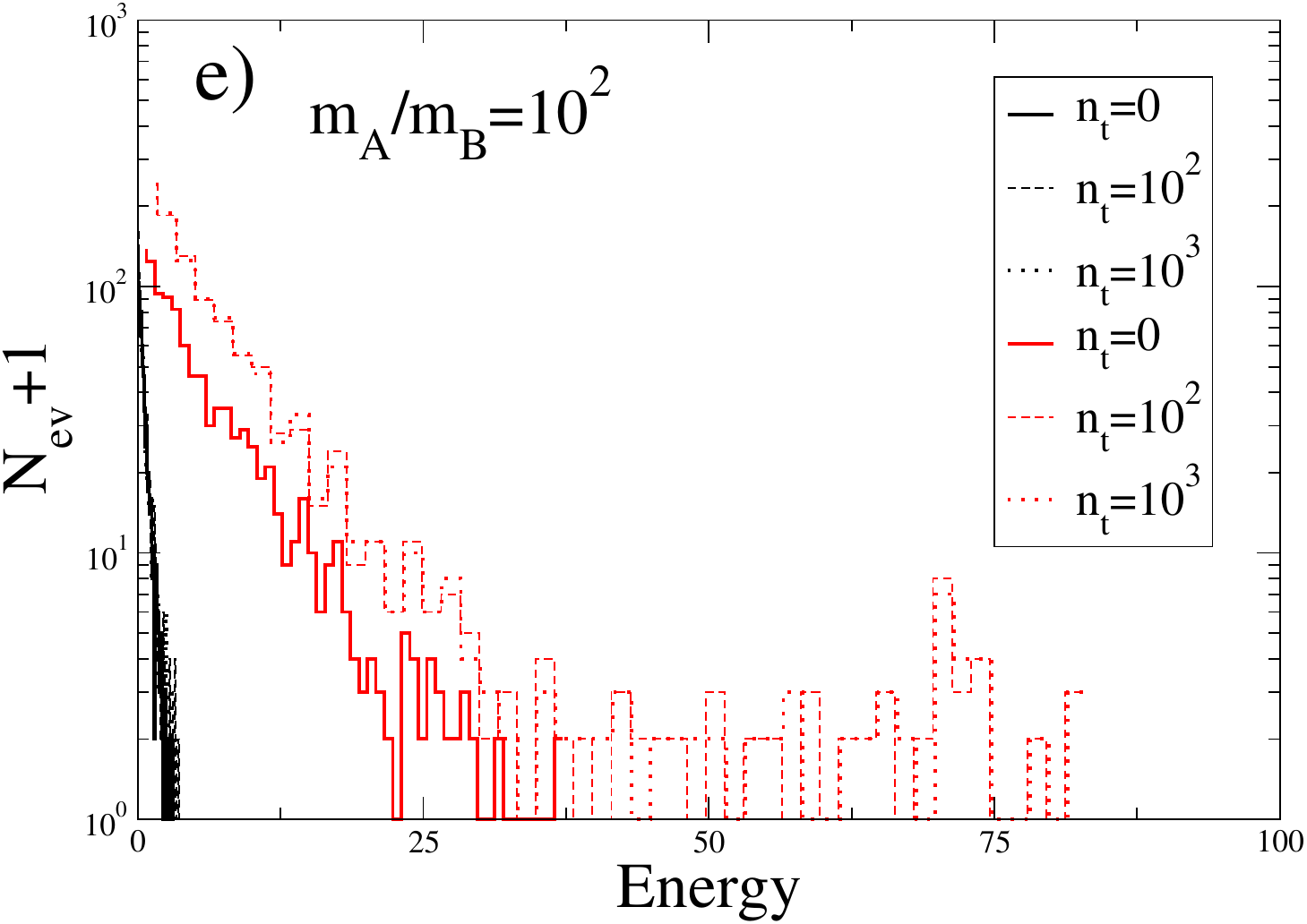}
\includegraphics[width=0.32\textwidth, clip=true]{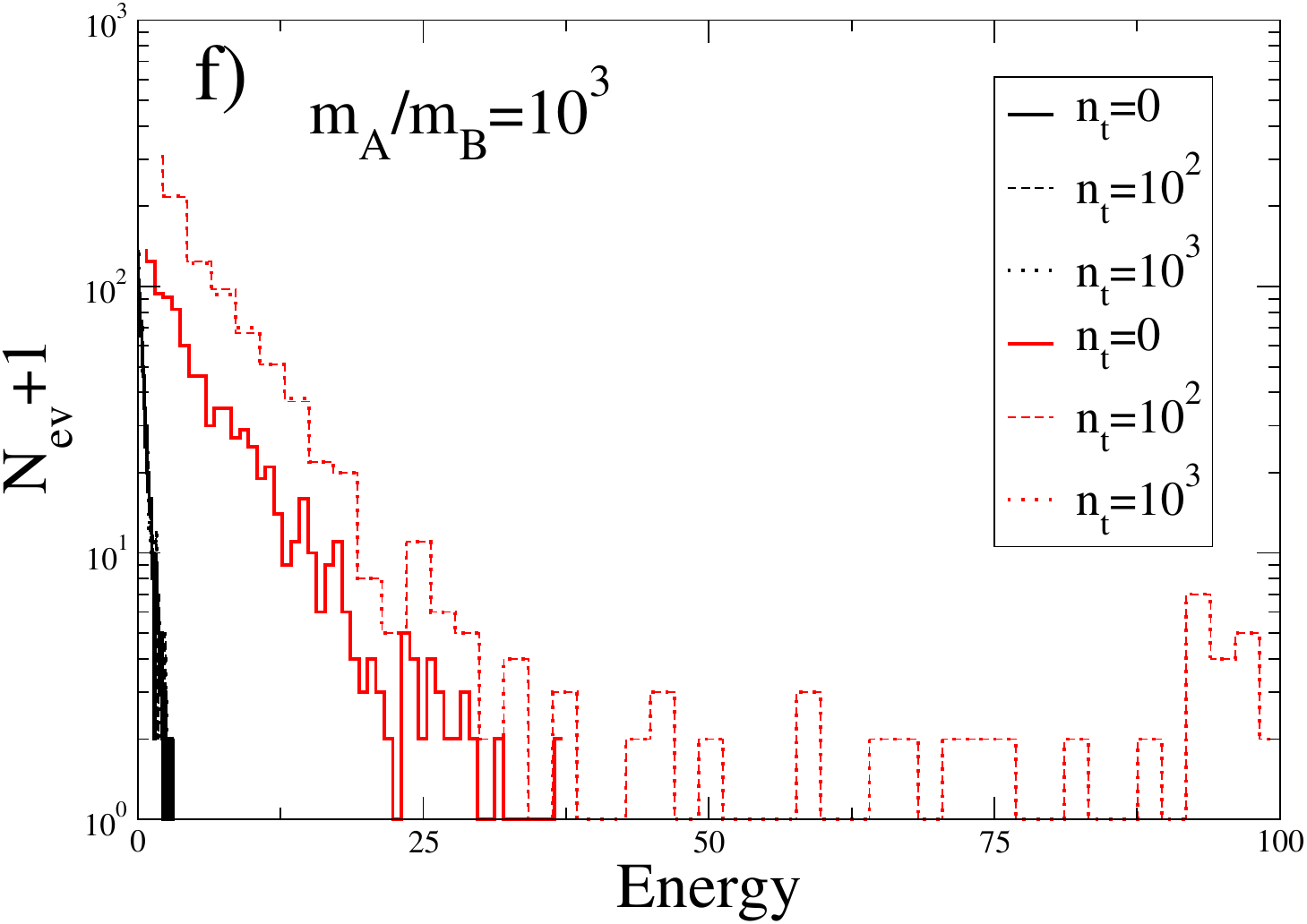}
\caption{Energy probability distributions ($N_{\mathrm{bins}}=50$) at initial time, after 100 times steps, and after 1,000 time steps for the 
two baths, and different mass ratios, for 1,000 oscillators in each bath, with $m_A$ indicating the mass of the coldest 
bath (black lines), and $m_A$ the hottest bath (red lines), all cases are for $\gamma=0.1$ and $\lambda=0.1$. 
To allow for an easy comparison we have adopted the same scales for all the plots.}
\label{Fig5}
\end{figure*} 
With this motivation, we assume the existence of a second MB distribution, at another inverse temperature for each of the baths $A$ and $B$.
Now the energy distribution for each bath depends upon three independent 
parameters, the two inverse temperatures $\beta_i^{(1)}$ and $\beta_i^{(2)}$ ($i$=A, B), and the probability $p_i^{(1)}$ that a particle belongs to the MB energy distribution with inverse temperature $\beta_i^{(1)}$. Now, the average number of particles in bath $i$ at inverse temperature $\beta_i^{(1)}$ is 
then $N_i^{(1)}=p_1^{(1)} N_i$. Similarly,  
the probability $p_i^{(2)}$ that a particle belongs to the MB energy distribution with inverse temperature 
$\beta_i^{(2)}$, leads the average number of particles at inverse temperature $\beta_i^{(2)}$ is $N_i^{(2)}=p_2^{(2)} N_i$, with $p_i^{(1)}+p_i^{(2)}=1$.
This results in 
\begin{eqnarray}
\Delta p_i^{bMB}&=& p_i^{(1)} [e^{-\beta_i^{(1)} E}-e^{-\beta_i^{(1)}(E+\Delta E)}] \nonumber \\ &+&p_i^{(2)}  [e^{-\beta_i^{(2)} E}-e^{-\beta_i^{(2)}(E+\Delta E)}],
\label{bMBprob}
\end{eqnarray} 
requiring a three-parameter fit, instead of the single parameter fit needed for the MB energy distribution case. It is clear from the outset that the last  row in  Fig. \ref{Fig3}  
 cannot be fitted adequately  either by  MB or bMB energy distributions. Attempts to do so result in large values of $\tilde{\chi}^2$ and, most importantly, in a 
large variability of fitting parameters with comparable $\tilde{\chi}^2$, indicating a lack of robustness in the fitting procedure. We have therefore avoided 
consideration of these cases, all corresponding to long-range interactions, in the analysis reported in Table I. 
Table I shows the results of best fit of the energy distributions of Fig.~\ref{Fig3} evaluated after $10^3$ time steps with MB and 
bMB energy distributions as in Eqs.~\ref{MBprob} and \ref{bMBprob} respectively. The presence of a `second inverse temperature' $\beta_i^{(2)}$ very close to 
$\beta_i^{(1)}$, a value of $p_i^{(1)}$ close to unity, or a quite similar value between the $\tilde{\chi}^2$  for an MB and a bMB, all indicate that there is no significant 
gain in considering a bMB. 

A distinctive feature in Table I is the presence of a significantly better fitting with a bMB energy distribution in the case of $\lambda=10^{-2}, \gamma=10^{-2}$, 
Fig.~\ref{Fig3}(a), in the case of the initially higher temperature bath only. All indicators show that a significant bMB develops, the probability $p_1^{(B)}$ is close to 0.5, 
$\tilde{\chi}_B^2$ for the bMB is less than half than for the MB, $\beta_B^{(1)}$ is significantly different from $\beta_B^{(2)}$ and both these values are 
distinct from $\beta_B^{(2)}$. All this occurs while the lower temperature bath A continues to be adequately represented by a single MB energy distribution.
This aspect is easily taken into account by considering the inhomogeneous nature of the cooling process. The species A at lower temperature is more populated 
at the center of the trapping potential, resulting in a large cooling power the minority species B at higher temperature in that region. By zooming in on the distributions of $\Delta p_A, \Delta p_B$ at low energy, as in  Fig.~\ref{Fig4}, we see clearly that the bath B has indeed a bimodal distribution, with a kink located around an energy of 
about 2.5. An analogous behavior, due to a mismatch in heat capacity, but only evident by eye due to the large statistical uncertainties at higher energy, seems 
appreciable for the bath A as the slope of the right part of the histogram seems compatible with a smaller inverse temperature than the left part, always with 
a putative kink in the same region as for bath B. We note that these features are evident when we use $10^4$ particles in each bath but not for
$10^3$ particles, as statistical fluctuations dominate and wash out the effects. 
While deviations from an MB energy distribution attributable to interaction strength and range are rather independent upon the spatial location of the two baths, this effect 
strongly depends upon the local density, in particular, the overlap of the two species. This feature can be further elucidated by considering an aspect we have
not explored earlier, namely baths with particles having different mass and/or trapping frequencies.
\begin{figure*}[t]
\includegraphics[width=0.32\textwidth, clip=true]{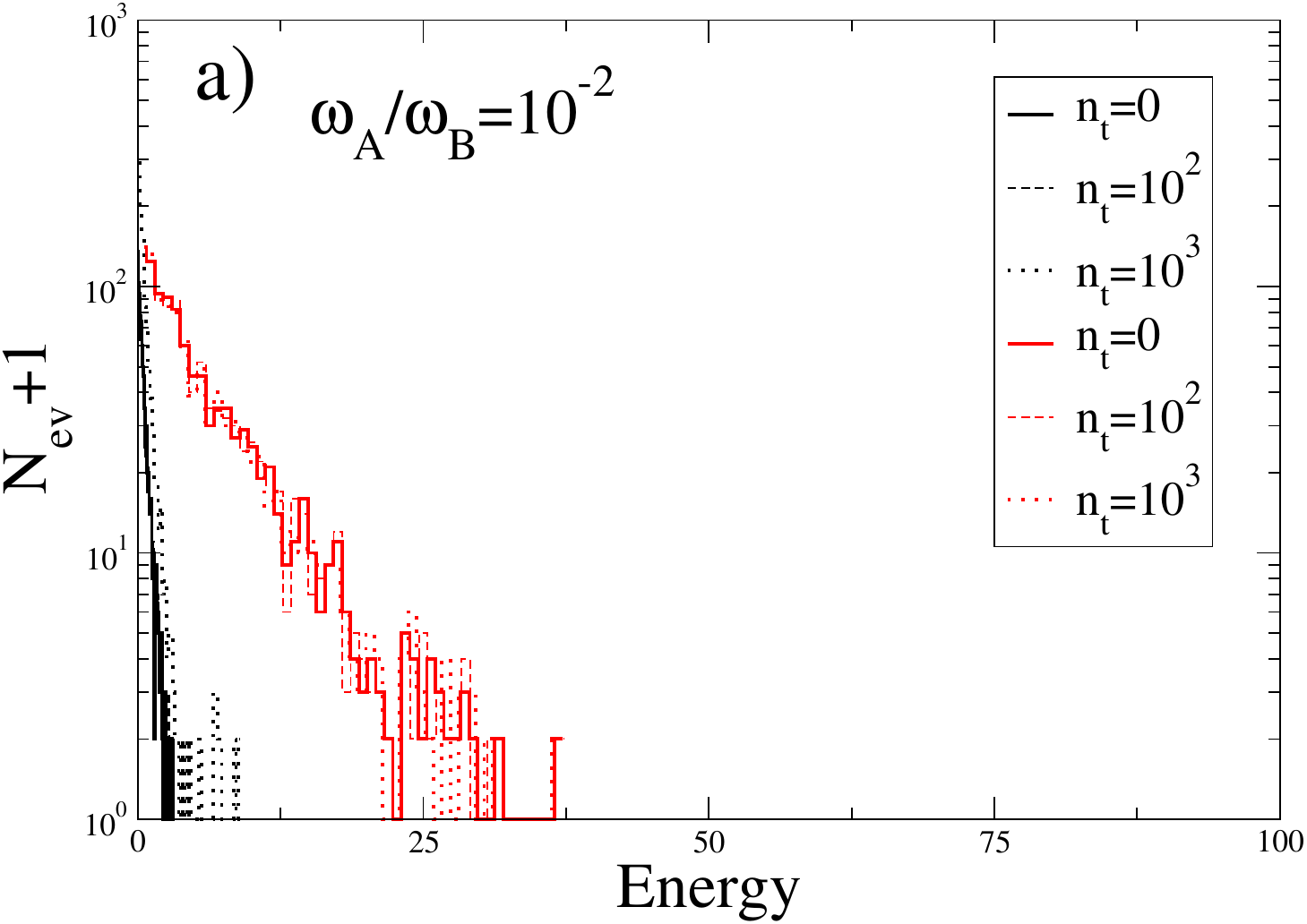}
\includegraphics[width=0.32\textwidth, clip=true]{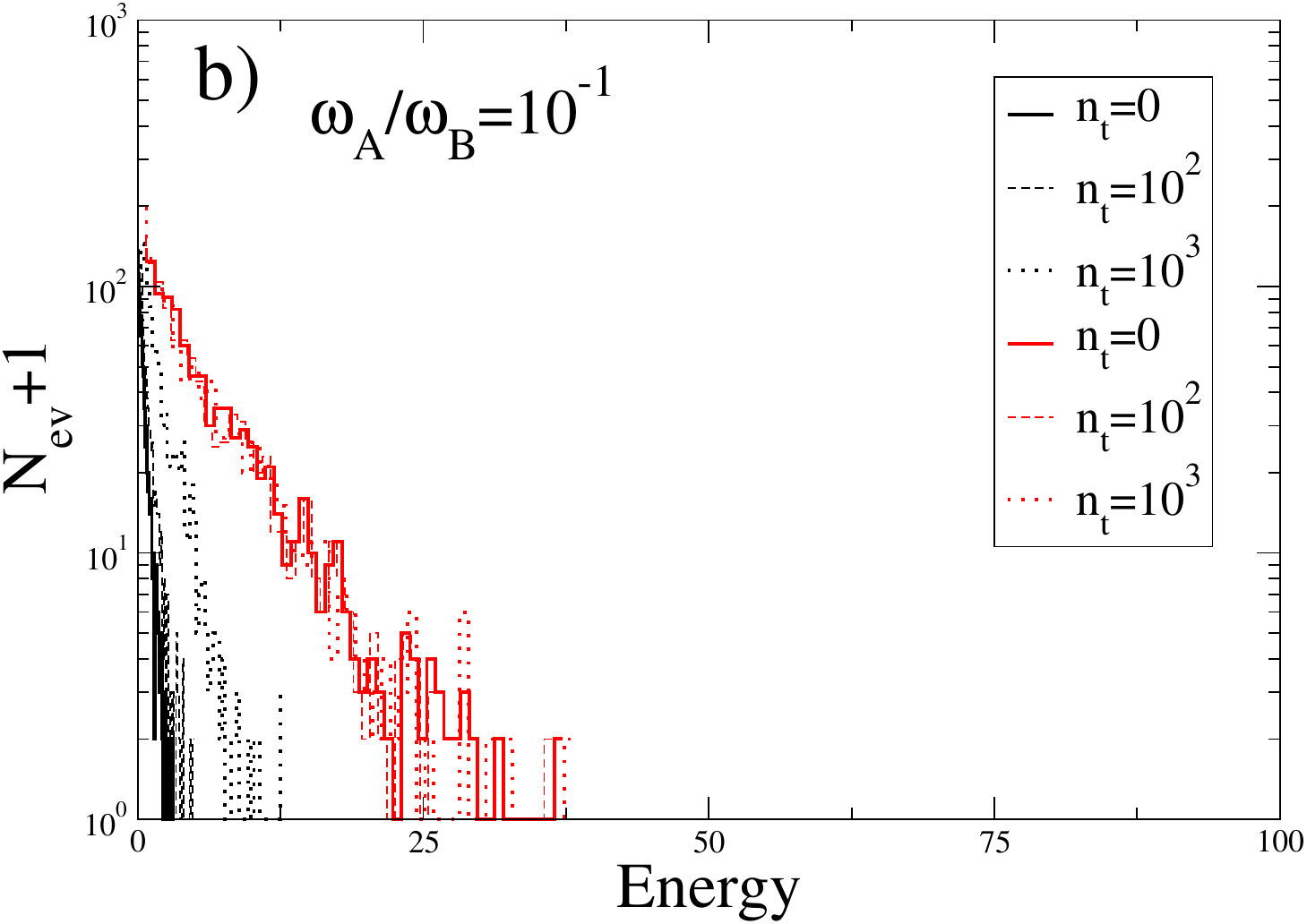}
\includegraphics[width=0.32\textwidth, clip=true]{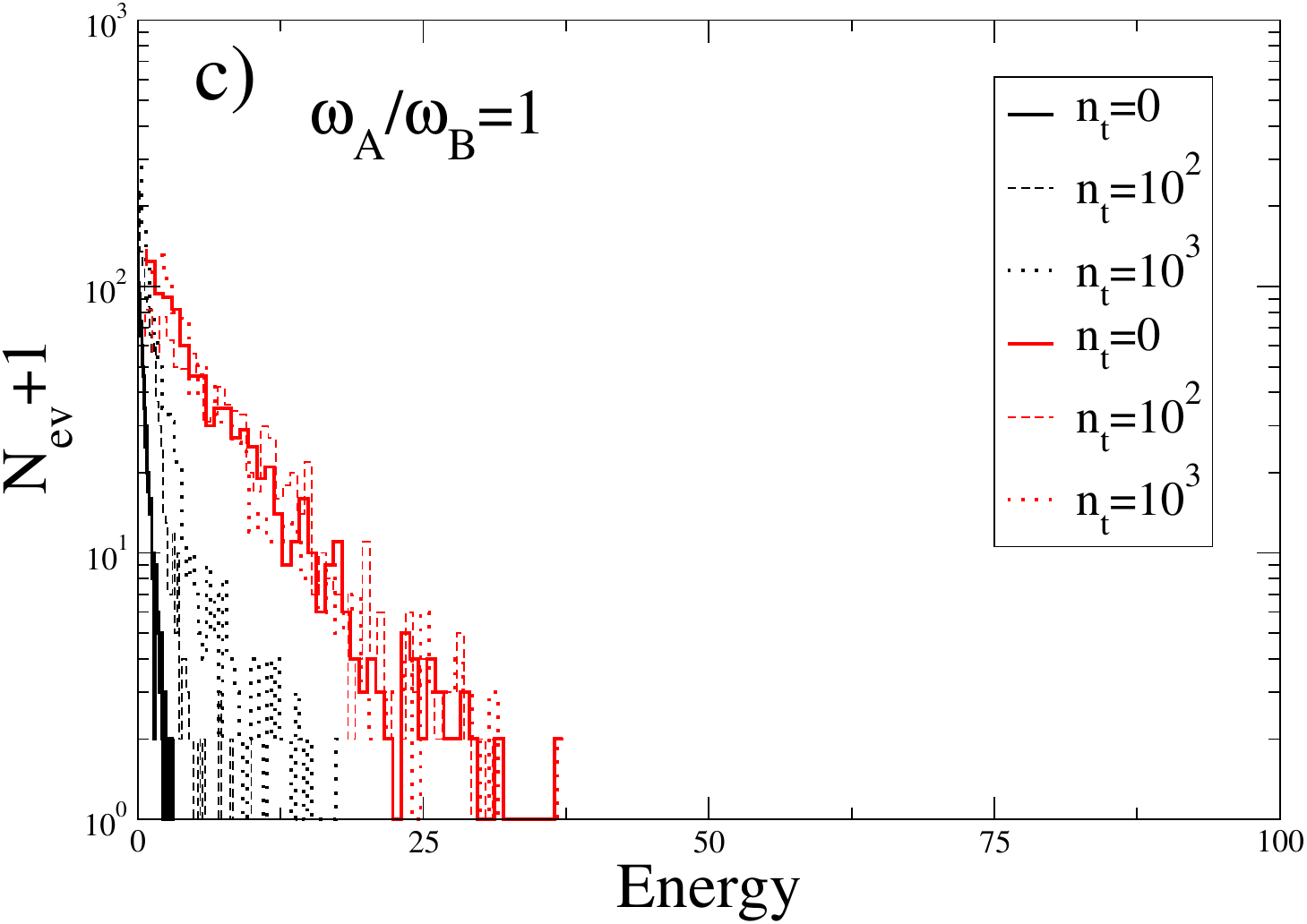}
\includegraphics[width=0.32\textwidth, clip=true]{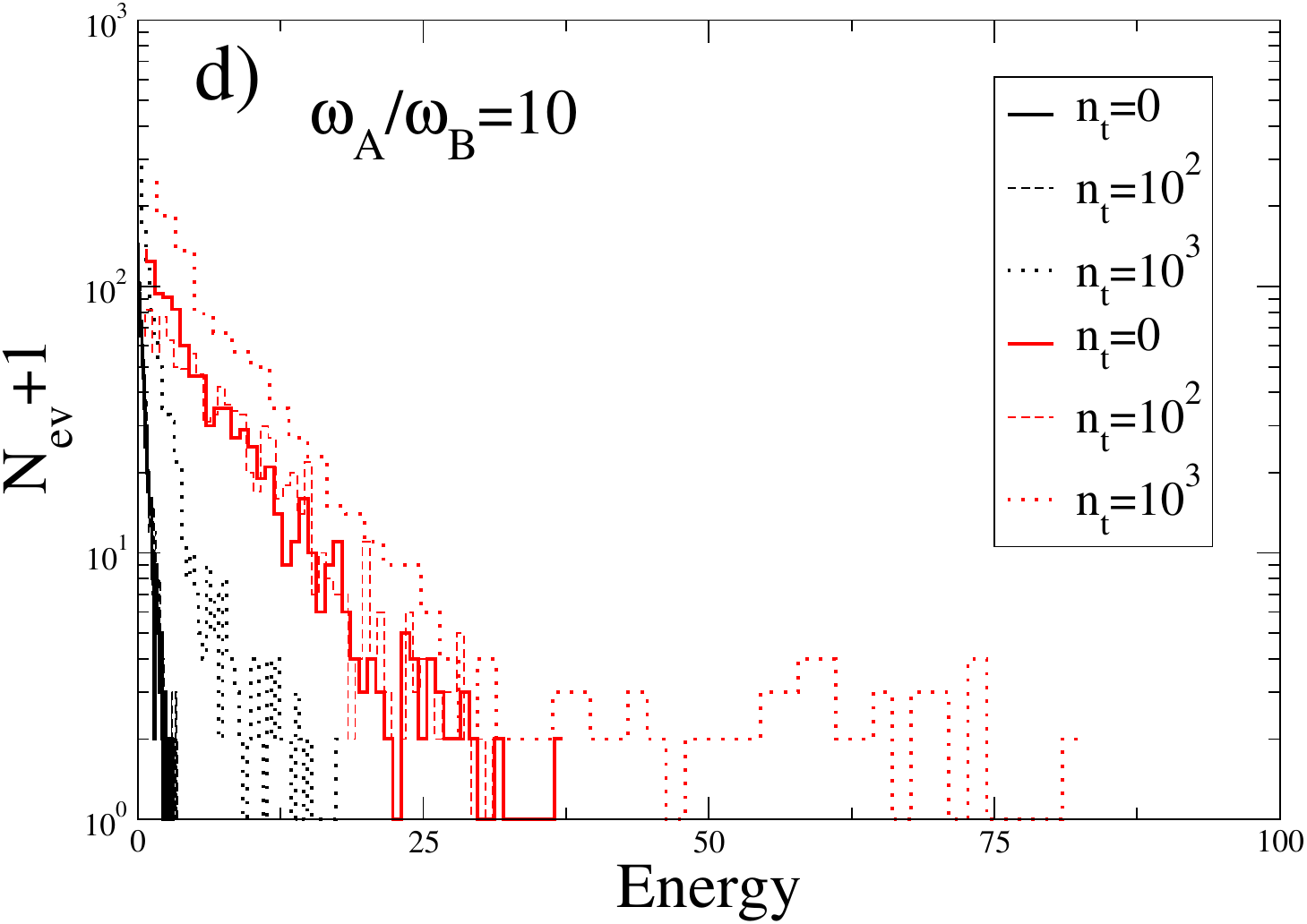}
\includegraphics[width=0.32\textwidth, clip=true]{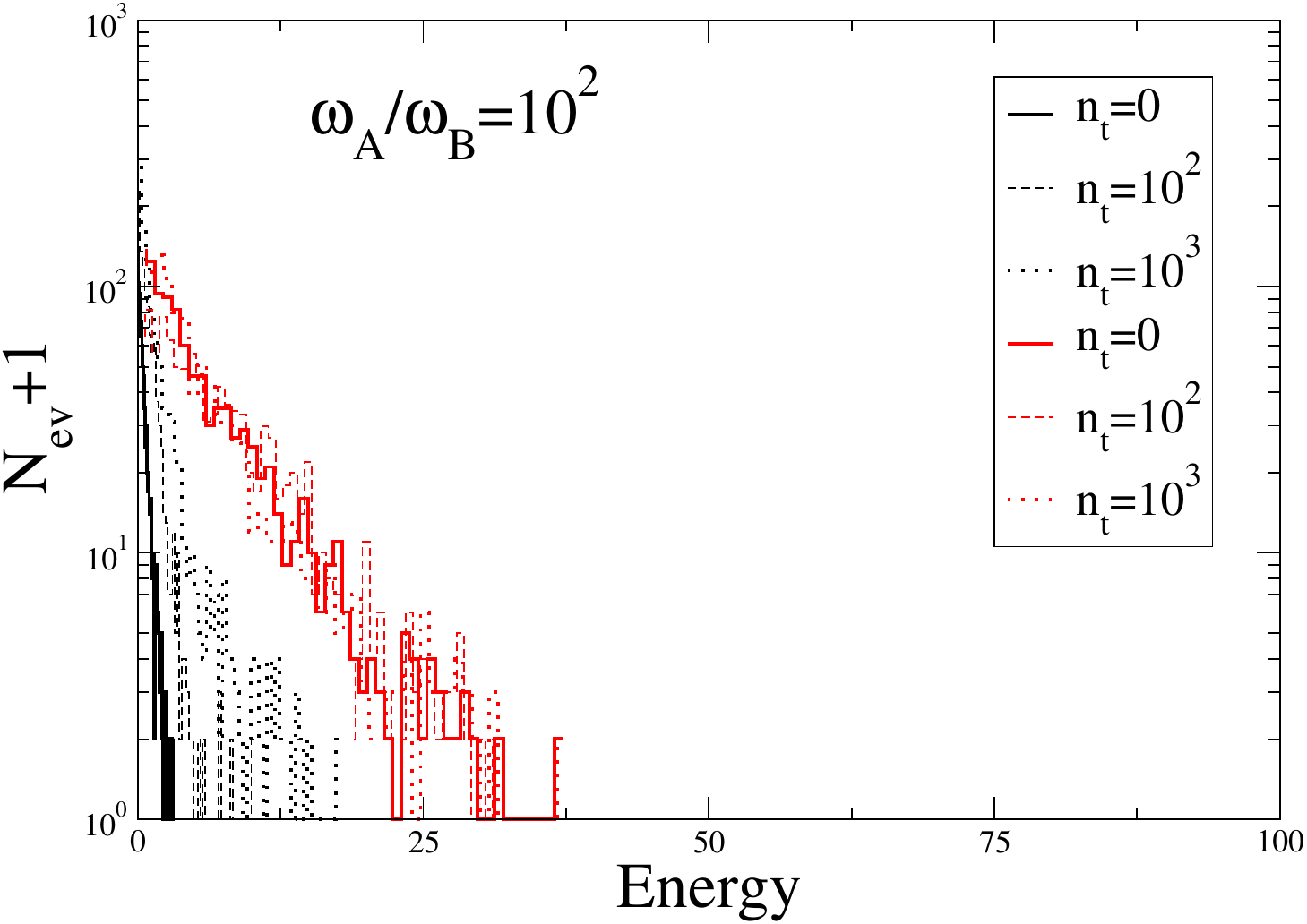}
\includegraphics[width=0.32\textwidth, clip=true]{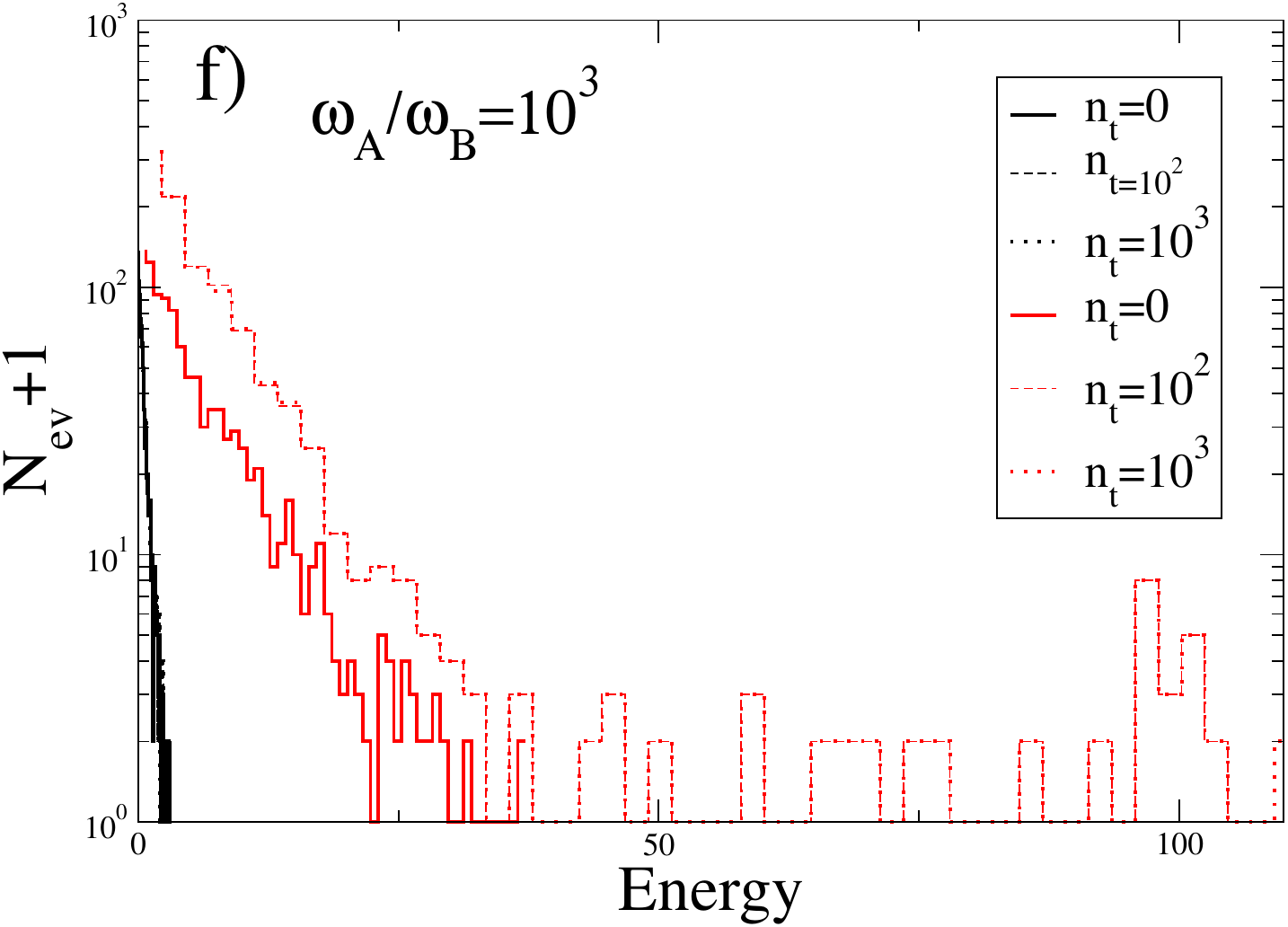}
\caption{Energy probability distributions ($N_{\mathrm{bins}}=50$) at initial time, after 100 times steps, and after 1,000 time steps for the 
two baths, and different frequency ratios, for 1,000 oscillators in each bath, with $\omega_A$ indicating the angular frequency of the coldest
bath (black lines), and $\omega_B$ the hottest bath (red lines), all cases are for $\gamma=0.1$ and $\lambda=0.1$ and equal mass $m_A=m_B$.}
\label{Fig6}
\end{figure*}
\section{Dependence of thermalization upon mass and trapping angular frequency of the particles}
Thus far, we have only considered two baths composed of particles with the same mass. From the standpoint of potential applications, it 
 is important to study the case of baths with particles 
having significant mass mismatch. 
For instance, in the case of a neutral plasma of electrons and ions \cite{Brown}, in cold mixtures of two different 
atomic species in magnetic and optical dipole traps \cite{Delannoy,Presilla}, and in experiments involving energy relaxation of fast atoms sent into a thermal bath gas, such as in \cite{Dalgarno}. The reaction times of the two baths, taken together with the fact that the heavier mass is basically unaffected in a single encounter with one from the
lighter species, can drastically alter the dynamics of thermalization. Historically, this case was first discussed by Lorentz, and 
it is therefore named after him. This study is especially important in relation to kappa energy distributions, as it has been argued 
\cite{Shizgal1,Shizgal2,Shizgal3,Shizgal4} that they only emerge in the limit of the mass ratio going to zero.

Our simulations showing the dependence on the mass ratio, at constant and equal trapping frequencies, are  seen in Fig.~\ref{Fig5}. 
We expect that a mass ratio significantly 
different from unity will affect the efficiency of thermalization as the more massive species will absorb or release very little energy and 
momentum during the interspecies interactions. This is visible in panel (a) of Fig.~\ref{Fig5} in which the high temperature bath has particles with a mass 
$10^2$, and reciprocally in panel (e). 
These cases share a common trend, as the energy distributions for the massive particles are nearly constant in time. 
As particularly visible in panel (f), the high temperature bath made of lighter particles 
develops a long high energy tail. This results from occasional interactions with the more massive particles resulting in additional energy increase on top of 
the release of the interaction potential energy from initial time, proportional to the interaction strength $\gamma$. 
In this dynamics, we also need to take into account that the mass of the particles also plays a role in the trapping potential, therefore the more massive particles  
experience a stronger confinement. This goes against a fast thermalization process, due to the diminished spatial overlap of the two baths. 
As expected, in the case of a mass ratio closer to unity, both the energy transfer at the microscopic level, as well as the spatial overlap, are maximized, resulting 
in a faster approach to thermalization.

An analogous study is possible with regard to the trapping frequencies. This differs from the previous one as the trapping frequencies 
appear only in the harmonic potential term, while the masses appear in kinetic term as well. For non-interacting 
systems the trapping frequencies determine the extent of the region in which most of the particles are located around the equilibrium 
point. A difference in their trapping frequencies, with the mass ratio kept at unity, therefore affects the overlap between the two species.
In the one-dimensional system analyzed here, it has to be kept in mind that particles encounter each other in any case (unlike 
higher dimensionality in which particles can avoid each other), however a very low frequency for one of the species  results
in significant residence in the outer region where the probability of encountering a particle of the other bath is negligible.

Ultimately, what matters for the energy exchange between the two bath is also the spatial overlap between them. In the simplest case 
of weak interspecies interactions, the spatial profile of each gas is given by a Gaussian dependence
\begin{equation}
\rho_i(q_i) \propto e^{-\beta_i m_i\omega_i^2 q_i^2/2},
\end{equation} 
therefore the spatial extension is given by the characteristic length $(\beta_i m_i \omega_i^2/2)^{-1/2}$, and the optimal overlap is achieved 
when this quantity is the same for the two gases. If the colder gas has also a small spatial extension (which happens if the mass or the angular frequency, or both, 
are very large with respect to the ones of the hotter gas), it will interact with a minute portion of the hotter gas, and strong deviations from MB are expected, as 
visible in the latest plots in Figs.~\ref{Fig5} and \ref{Fig6}. 

This aspect is relevant at least in two contexts. In ultracold atom physics, two species simultaneously trapped can efficiently thermalize 
depending on their mass ratio and frequency ratio. In a magnetic trap for instance, the magnetic moments of 
alkali metal atoms are rather similar, so the magnetic force acting on atoms is comparable for different species. Then the 
product of the mass times the square of the trapping frequencies is approximately constant, therefore the species with the 
larger mass will oscillate with a longer period. In optical dipole traps the situation can be rather complex instead, and rather arbitrary 
values of trapping frequencies can also be achieved by using species-selective trapping as in bichromatic setups. 
Another subfield in which these considerations are relevant is the confinement of electrons and ions, for instance in neutral plasmas,
for controlled nuclear fusion. In the presence of mass-independent forces like the Lorentz force, the electrons will have a faster dynamics 
due to their mass, but at the same time the transfer of heating to the ions will be quite inefficient. Of course, the considerations reported here have 
partial validity in this case because of the assumption of negligible intra-species interactions, which is manifestly inaccurate in the case 
of fully ionized plasmas.

\section{Number-Unbalanced baths} 

We finally consider the case of baths in which $N_A \neq N_B$, with particular emphasis on on the case of large imbalance, {\it i.e.} $N_A \gg N_B$. 
This is relevant for Bose-Fermi mixtures of cold atoms and in the case of charged plasmas. In the case of strongly interacting baths, the interaction energy 
contributes to the equilibrium temperature. We recall, from the discussion detailed in \cite{JauOnoSun1}, that an equilibrium temperature can then be 
assigned after thermalization, based purely on conservation of the total energy, as
\begin{equation}
T_{\mathrm{eq}}=\frac{N_A T_A + N_B T_B}{N_A+N_B} + \frac{\langle E_{\mathrm{int}}\rangle}{(N_A+N_B)k_B},
\label{eqtemp}
\end{equation}
where $k_B T_{\mathrm{eq}}=(\beta_{\mathrm{eq}})^{-1}$, $k_B T_A=(\beta_A)^{-1}$, $k_B T_B=(\beta_B)^{-1}$, and 
$E_{\mathrm{int}}$ is the value of the last term in Eq.~\ref{Hamilton} at equilibrium.
Since $E_{\mathrm{int}} \propto \gamma$, the equilibrium temperature is intermediate between the two initial temperatures in the 
weak coupling limit, as expected by considering the heat capacities of the two baths in a thermal setting. However, in the strong 
coupling limit $T_{\mathrm{eq}}$ can be larger than the temperature of the hottest bath. 
Notice also that, since in general $\langle  E_{\mathrm{int}}\rangle$ is not extensive, $T_{\mathrm{eq}}$ is not an intensive quantity.

It is possible to evaluate the correction to the equilibrium temperature in the weak coupling regime. The last term of the right hand side of 
Eq.~\ref{Hamilton1} can be used to define the average value of its exponential term as
\begin{equation} 
\langle  e^{-(Q-q)^2/\lambda^2} \rangle = 
\frac{1}{N_A N_B}\sum_{m=1}^{N_A} \sum_{n=1}^{N_B} e^{-(Q_m-q_n)^2/\lambda^2},
\end{equation}
and, using the cumulant approximation 
\begin{equation}
\langle e^{f(x)} \rangle \simeq e^{\langle f(x) \rangle},
\end{equation}
after few steps, by averaging on an MB energy distribution, we obtain a self-consistent relationship for the equilibrium temperature
\begin{equation}
T_{\mathrm{eq}} \simeq 
\frac{N_A T_A+N_B T_B}{N_A+N_B}+ \frac{\gamma}{k_B} \frac{N_A N_B}{N_A+N_B} 
e^{-\frac{k_B T_{\mathrm{eq}}}{K_{\mathrm{eff}} \lambda^2}},  
\label{selfconst}
\end{equation}
where we have introduced an effective combined stiffness constant $K_{\mathrm{eff}}$, always positive definite, as
\begin{equation}
K_{\mathrm{eff}}^{-1}=\frac{2}{m_A \omega_A^2}+\frac{2}{m_B \omega_B^2}-\frac{\pi}{\sqrt{m_A m_B} \omega_A \omega_B}.
\end{equation}
Again, for small $|\gamma|$, $T_{\mathrm{eq}}$ in the last term of the right hand side of Eq.~\ref{selfconst} can be approximated as 
$T_{\mathrm{eq}} \simeq  (N_A T_A+N_B T_B)/(N_A+N_B)$,  together with the leading order correction to the equilibrium temperature.  
Eq. \ref{selfconst} shows that the $\gamma$-dependent term has a nonlinear dependence on the ratio between the amplitude of the motion 
of a particle with mean energy $k_B T$ and the range of interaction $\lambda$. In the limit of very large amplitude of the motion, 
$(k_B T/K_{\mathrm{eff}})^{1/2} \gg \lambda$, the interaction between particles is negligible as the time spent within a distance $\lambda$ is negligible, 
and the fact  that the particles will spend most of the time near the extrema of the motion, {\it i.e.} in low-density regions. 
Notice that the $\gamma$-dependent term in the unbalanced limit, $N_A \gg N_B$ or $N_A \ll N_B$, becomes linear in $N_A$ or $N_B$ respectively;
while in a more balanced situation $N_A \simeq N_B$ it is non-extensive, as manifest in Eq.~\ref{selfconst}. In any event, the equilibrium temperature is 
not an intensive quantity, and may become larger than each of the initial temperatures $T_A$, $T_B$ as $\gamma N$ exceeds a threshold value. However
the bistability mentioned earlier limits the maximum value of $\gamma$ for which this analysis is valid. 
Analogous considerations may be repeated in the case of interspecies attractive interactions, $\gamma < 0$, potentially leading to temperatures 
smaller than the initial ones. The possibility of having negative temperatures for $\gamma$ negative and large enough in absolute value has to be interpreted, in this framework, 
as the presence of instability for the two baths, leading to the collapse of the two clouds in the common minimum of the trapping potential.

For completeness, it is also worth considering the particular case of two baths with equal parameters, $N_A=N_B=N/2$, $m_A=m_B=m$, $\omega_A=\omega_B=\omega$, as we obtain a rather simple expression, valid  in the $|\gamma|\ll 1$ limit
\begin{equation}
T_{\mathrm{eq}} \simeq \frac{T_A+T_B}{2}+ \frac{\gamma N}{4k_B} \exp\left[-\frac{4-\pi}{2} \frac{k_B (T_A+T_B)}{m \omega^2 \lambda^2}\right],
\label{Balanced} 
\end{equation}
which clearly shows the correction to the equilibrium temperature, expected for a very weak interaction linear in $\gamma$, and non-perturbative in $\lambda$.
The latter dependence leads to the correction quickly going to zero in the $\lambda \to 0$ limit of pointlike interactions regardless of $\gamma$, and reaching the 
maximum value of $\gamma N/(4 k_B)$ in the $\lambda \to \infty$ limit. It should be noted that the estimate of the correction to the equilibrium temperature 
relies on evaluating the averages with 
an ensemble approach. This in turn relies on the validity of the ergodic hypothesis, which is not necessarily enforced in the case of 
long-range interactions, yet another nuance we plan to discuss in a future contribution.

Finally, we recall that in \cite{JauOnoSun1} we had identified intriguing behavior that 
the oscillation frequency of the minority species, in the strongly interacting limit, becomes locked to the 
frequency of the majority species. This effect also depends on the encounter between particles of the two species needed for thermalization, and 
therefore we expect it to be proportional to the coupling strength $\gamma$ and the range $\lambda$ in a similar fashion. 
\section{Conclusions}
Using numerical simulations, we have discussed the dynamics of thermalization, with particular emphasis on deviations from the MB energy 
distributions, and its dependence upon the parameters of a model in which two baths of harmonically confined particles interact with a Gaussian 
dependence on the distance. This study shows explicitly that the quasi-static expectation of maintaining an MB energy distribution en route to thermalization
is more the exception than the rule, and occurs only for
a restricted range of the parameters characterizing the interaction between the two thermalizing systems.
If the interaction range is small, we expect validity of the Boltzmann assumptions on molecular chaos. However the interactions with significant 
exchange of energy are rare, therefore thermalization occurs on long time scales, even more so if the interaction strength is also small. 
At intermediate times, thermalization is partial and mainly occurs for the low energy component, as expected 
 from the fact that particles closer to the minimum of the trapping potential have on average a larger interaction rate. 
More energetic particles,  that spend most of their time at the periphery of the potential have less frequent interactions. 
Of course, by increasing the interaction range,  larger numbers of interactions occur at a given time. In the limit of 
a range much larger than the average separation of two particles we fall into the Caldeira-Leggett regime, and unless the 
two baths have the same frequency, we do not expect thermalization. 

We have identified that deviations from MB energy distributions occur on the road to thermalization due to three independent sources, Specifically,
strong coupling, intermediate interaction range, and also due to the inhomogeneous nature of the trapping potential. 
Computational limitations in our simulations prevent us from definitively identifying the non MB energy distribution as a kappa distribution, 
as expected in various contexts. However,  we have discussed the possibility of fitting the data with a bMB distribution instead, with which 
a kappa distribution can be efficiently approximated \cite{Hahn}.
This minimalistic framework of bMB energy distributions also allows for a conceptual understanding of deviations in the intermediate 
interaction range, as in this case a bistable dynamics is a precursor of the onset of chaos-  as seen in earlier work~ \cite{OnoSun2}. 
This corresponds to a situation where each bath interacts with two effective baths maintained at different temperatures.  An additional intriguing inference is
that, for specific parameter domains, the thermalization process could transit through a sequence of reducing multi-temperature MB distributions 
before arriving at a bMB and, finally, a single MB. This would suggest a reverse bifurcation route \cite{Cascais} from dynamics to thermodynamics.

The dependence on the intrinsic parameters of each bath, {\it i.e.} mass and the trapping angular frequency of the particles, and their number, has also been 
discussed, with emphasis on the loss of efficiency in the thermalization of particles differing in mass and/or trapping angular frequencies, as well as 
intriguing aspects of baths with number unbalance. In particular, the mass dependence is relevant for heating of a more massive species via 
contact with a much lighter species, as happens in the case of confined plasmas used for nuclear fusion. In this specific context, it is relevant to 
remark that non MB energy distributions have been observed, as in \cite{Bhatnagar,Kovtun}, and are expected to play a major role in self-burning plasmas \cite{Kiptily}.

\end{document}